\documentclass[
 reprint,
 amsmath,amssymb,fleqn,
 aip,
 jcp,
]{revtex4-1}

\usepackage[english]{babel}
\usepackage[utf8x]{inputenc}
\usepackage{amssymb}
\usepackage{amsmath}
\usepackage{amsfonts}
\usepackage[bbgreekl]{mathbbol}
\usepackage{graphicx}
\usepackage{color}
\usepackage{hyperref}

\newcommand{\bra}{\langle}
\newcommand{\ket}{\rangle}

\newcommand{\qq}{\mathbf{q}}
\newcommand{\dd}{\mathrm{d}}
\newcommand{\ee}{\mathrm{e}}
\newcommand{\Pro}{\mathrm{pr}}
\newcommand{\OmegaR}{\Omega_\mathrm{R}}

\newcommand{\kk}{\mathbf{k}}
\newcommand{\rr}{\mathbf{r}}
\newcommand{\dg}{\dagger}

\begin{document}

\title{Quantum dynamics and spectroscopy of dihalogens in solid matrices.\\
II.\ Theoretical aspects and G-MCTDH simulations of time-resolved coherent Raman spectra of
Schr\"odinger cat states of the embedded $\mathrm{I_2 Kr_{18}}$ cluster.}  
\author{David Picconi}
\email{picconi@chemie.uni-frankfurt.de}
\affiliation{Institute of Physical and Theoretical Chemistry, Goethe University Frankfurt, Max-von-Laue-Stra{\ss}e 7, D-60438 Frankfurt am Main, Germany}
\author{Jeffrey A. Cina}
\affiliation{Department of Chemistry and Biochemistry, and Oregon Center for Optical, Molecular, and Quantum Science, University of Oregon, Eugene, Oregon 97403, USA}
\author{Irene Burghardt}
\affiliation{Institute of Physical and Theoretical Chemistry, Goethe University Frankfurt, Max-von-Laue-Stra{\ss}e 7, D-60438 Frankfurt am Main, Germany}

\date{\today}

\begin{abstract}
This companion paper to [D. Picconi et al., J. Chem. Phys. 150 (2019)] presents quantum dynamical simulations, using the Gaussian-based
multiconfigurational time-dependent Hartree (G-MCTDH) method, of time-resolved
coherent Raman four-wave-mixing spectroscopic experiments for the iodine molecule embedded in a
cryogenic crystal krypton matrix. 
These experiments monitor the time-evolving vibrational coherence between two wave packets created in a quantum superposition (i. e. a \lq Schr\"{o}dinger cat state') by a pair of pump pulses which induce electronic $B\ ^3\Pi_u\left(0^+\right) \longleftarrow X\ ^1\Sigma_g^+$ transitions.
 A theoretical description of the spectroscopic measurement is developed, which
elucidates the connection between the nonlinear signals and the wave packet
coherence. The analysis provides an effective means to simulate the spectra for several different optical conditions with a
minimum number of quantum dynamical propagations. The G-MCTDH method is used
to calculate and interpret the time-resolved coherent Raman spectra of two
selected initial superpositions for a $\mathrm{I_2 Kr_{18}}$ cluster embedded in a frozen Kr cage.
The time- and frequency-dependent signals carry information about the molecular mechanisms of dissipation and decoherence, which involve vibrational energy transfer to the stretching mode of the four \lq belt' Kr atoms. The details of these processes and the number of active solvent modes depend in a
non-trivial way on the specific initial superposition.
\end{abstract}

\maketitle

\section{Introduction}

Halogen molecules embedded in rare gas crystals have been often regarded as prototypical systems for the investigation of condensed phase dynamics.\cite{AS99,GBFKS07,kuehn-woeste-book} The molecular bond elongation, initiated by an electronic excitation, leads to a collision with the surrounding matrix, and the ensuing solute-solvent interactions can be investigated in great detail -- given the simplicity of the chromophore -- using today's sophisticated techniques of nonlinear spectroscopy.\cite{mukamel} Moreover, the large number of experimental data are beneficial to the development of new theoretical methods for simulating the photophysics and photochemistry of embedded chromophores\cite{PCB18A,BC97,OA98,BK07,BGGSHJ12,BGC18} and for calculating the related nonlinear spectroscopic signals.\cite{GED09,KFDBM17} 

\begin{figure}[b!]
\centering
\includegraphics[scale=0.33]{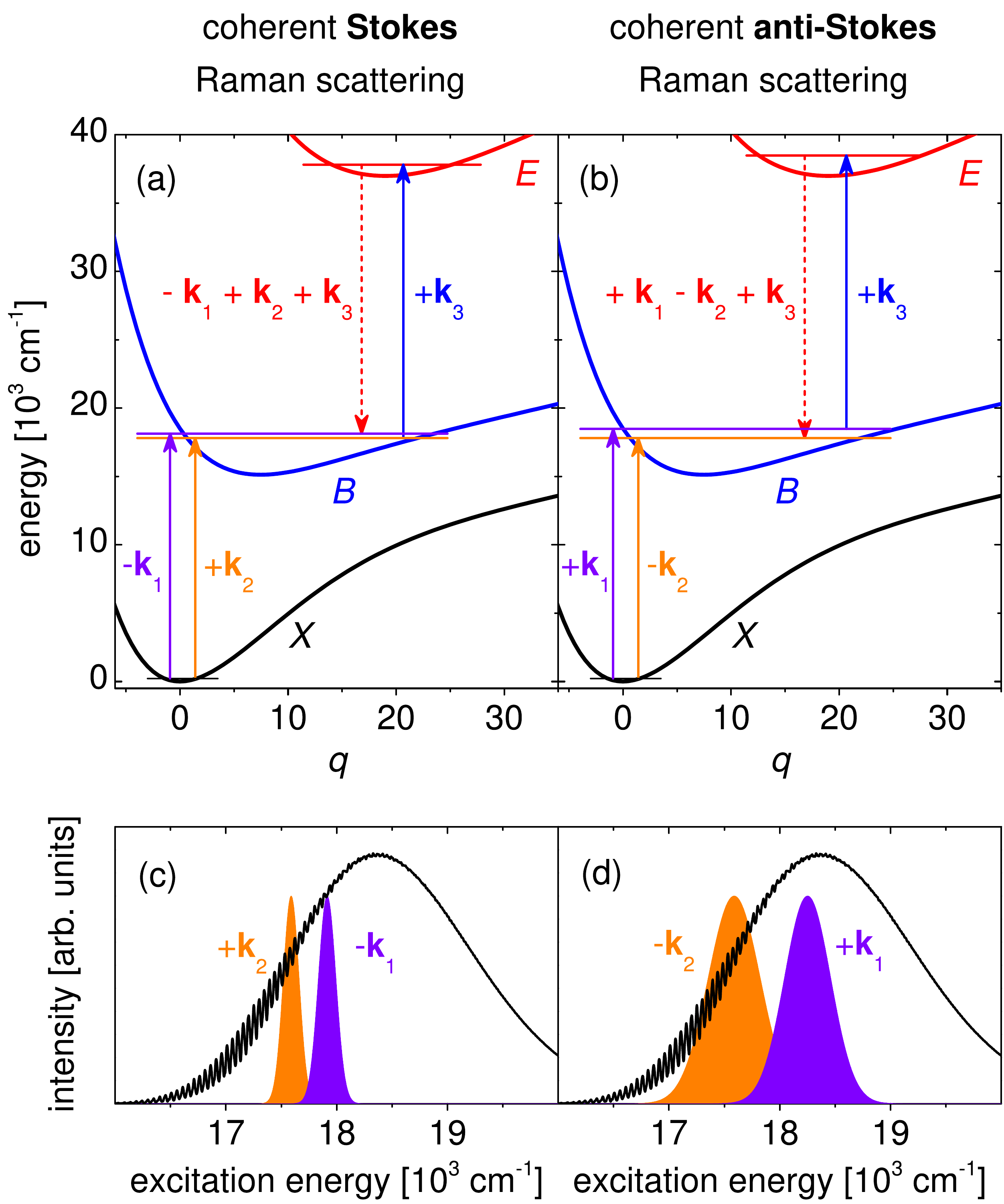}
\caption{(a,b) One-dimensional potential energy cuts for the electronic states $X \ ^1\Sigma_g^+$, $B \ ^3\Pi_u(0^+)$ and $E (0_g^+)$ of molecular iodine embedded in crystal krypton along the dimensionless normal mode $q$ associated to the I--I bond stretch. The purple and orange arrow signify a pair of pump pulses tagged by the wavevectors $\pm\mathbf{k}_1$ and $\pm\mathbf{k}_2$, which create a wave packet superposition on the $B$ surface. The blue and red arrows indicate the $E \longleftarrow B$ transition induced by the probe pulse, and the subsequent coherent spontaneous $B \longleftarrow E$ emission, which is detected in the Stokes (a) and anti-Stokes (b) directions. The magnitudes of the arrows in panels (a) and (b) refer to the settings of the calculations \textbf{A} and \textbf{B} of Sect. \ref{sec: calculated_signal}, respectively. (c,d) The linear absorption spectrum (black lines) for the $B \longleftarrow X$ transition calculated in Ref. \onlinecite{PCB18A} using G-MCTDH quantum dynamics; the orange and purple profiles in panels (c) and (d) depict the power spectra of the laser pulses used in calculations \textbf{A} and \textbf{B}, respectively.} 
\label{fig: Pulses}
\end{figure}

In particular, the photodynamics of molecular iodine in solid krypton has been studied with
several nonlinear spectroscopy
techniques,\cite{BCWYAMZKM97,BDDS99,AKZA00,BGDS02,SKFA05,SA11} which can be classified into two main classes. The first class includes pump-probe spectroscopies in which a first pulse prepares a wave packet in an excited electronic state and a second pulse is used to probe the time evolution of the reduced density matrix of the $\mathrm{I_2}$ chromophore.\cite{ZSA96,BC97,GBS03,BGS04} In this way, the dissipation of energy to the Kr environment can be monitored in time for different pump energies and effective molecular potentials can be reconstructed.\cite{BDDS99,BGDS02,GBFKS07} The second class of nonlinear spectroscopies specifically addresses quantum mechanical effects in solute-solvent interactions and makes use of four-wave-mixing optics.\cite{SKFA05,SA11} In the experiments performed by Apkarian and co-workers a \lq cat'-like superposition of two
wave packets, $\chi_1^B(t)$ and $\chi_2^B(t)$, is created on the $B \ ^3\Pi_u(0^+)$ state by interaction with a pair of excitation pulses. The quantum mechanical coherence between the wave packets is monitored by time-resolved resonance Raman scattering via the intermediate $E (0_g^+)$ state,
induced by a \lq probe' pulse. Such a process is illustrated in Figs. \ref{fig: Pulses}(a) and (b). The detected observable is the third-order polarization as a function of the pump-probe delay $T_\Pro$. 

Time-resolved coherent Raman spectroscopy therefore provides an ideal test-bed for the study of the entanglement and the transition to classicality in embedded molecular systems.\cite{SKFA05} The interpretation of third-order signals is however not always straightforward, so that theoretical methods are necessary to simulate the spectra and to correlate them to the underlying molecular dynamics.\cite{DS97,C08,BVM12} 

Four-wave-mixing experiments addressing vibrational coherences on the $X$ state, were simulated successfully using the semiclassical Liouville method;\cite{RFM06} the focus was on the coherences $\rho_{0n}$ between the vibrational ground state and the excited levels. For wave packet superpositions created in the electronically excited $B$ state, the coherence dynamics involve a much larger number of pairs of vibrational levels, so that hundreds of density matrix elements $\rho_{nm}$ should be propagated and an accurate semiclassical treatment becomes cumbersome. In addition, more rigorous, fully quantum, approaches are essential for the computation of full time- and frequency- resolved two-dimensional spectra.

Based upon the studies reported in the companion paper (henceforth referred to
as paper I),\cite{PCB18A} we now undertake to simulate the time-resolved coherent Raman spectra of $\mathrm{I_2}$ in crystal $\mathrm{Kr}$ using
the Gaussian-based multiconfigurational time-dependent Hartree (G-MCTDH) method.\cite{BMC99,BGW08}
As shown in paper I, the reduced subsystem density matrices of an $\mathrm{I_2
Kr_{18}}$ cluster calculated with the computationally inexpensive G-MCTDH approach are in
excellent agreement with the same quantities obtained at the
numerically exact multiconfigurational time-dependent Hartree (MCTDH) level.\cite{BJWM00}
This shows that G-MCTDH faithfully describes vibrational coherence and
time-evolving system-environment correlations. 
The method is therefore a highly valuable tool in the context of theoretical nonlinear spectroscopy and permits the efficient simulation of signals obtained for different pump-probe frequencies, pulse duration, pulse delays, etc.
It is therefore worthwhile, in order to fully illustrate the potential of
the G-MCTDH approach, to calculate a coherent nonlinear optical signal
which quantitatively monitors molecular entanglement and decoherence. 

Specifically, a theoretical analysis of the four-wave-mixing
experiments of Refs. \onlinecite{SKFA05} and \onlinecite{SA11} is developed to simulate the signals using
a minimum amount of quantum dynamical calculations, and to establish a
quantitative connection between the vibrational coherence and the spectra.
Time-resolved coherent Stokes and anti-Stokes Raman spectra of the $\mathrm{I_2
Kr_{18}}$ cluster are then calculated for different initial wave packet
pairs, which are created by $B\ ^3\Pi_u(0^+) \longleftarrow X \ ^1\Sigma_g^+$ transitions induced by a sequence of two
pump pulses. The transient resonance Raman probe transition to the $E(0_g^+)$ state is simulated, and the nontrivial features of the theoretical and experimental signals are then compared in detail.

The manuscript is organized as follows.
Sect. \ref{sec: theory} discusses the spectroscopic signal and the initial state
preparation, presents an analysis of the coherent Raman spectra and develops an approximate method to evaluate
this signal. In Sect. \ref{sec: comp. det.} computational details are given, and Sec.\ \ref{sec: calculated_signal} presents calculated signals for two different superposition states. Finally,
Sec.\ \ref{sec: Conclusion} summarizes the results and discusses future prospects.

\section{Time-resolved coherent Raman spectroscopy}
\label{sec: theory}
In the four-wave-mixing experiments of Apkarian and coworkers\cite{SKFA05,SA11} nonlinear polarization is induced in the $\mathrm{I_2:Kr}$ system by an interaction with a sequence of three femtosecond laser pulses. The pulses' wavevectors $\kk_1, \ \kk_2 \mbox{ and } \kk_3 $ are conventionally chosen such that $|\kk_1| > |\kk_2|$, and the emitted polarization is detected in the phase-matching directions $\kk_\mathrm{S} = - \kk_1 + \kk_2 + \kk_3 $, referred to as \lq Stokes direction' (coherent Stokes Raman scattering, CSRS), and $\kk_\mathrm{AS} = + \kk_1 - \kk_2 + \kk_3$, denoted \lq anti-Stokes' direction (coherent anti-Stokes Raman scattering, CARS).

We describe these experiments using a Hamiltonian which includes the electronic states $X$, $B$ and $E$ of the embedded iodine molecule,  which are depicted in Fig. \ref{fig: Pulses},
\begin{eqnarray}
\hat{H} & = & \hat{H}_0 + \hat{V}_\mathrm{int} \nonumber \\
 & = & \sum_{\alpha = X,B,E} |\alpha\ket \hat{H}_\alpha \bra \alpha | + \hat{V}_\mathrm{int} \ ,
\end{eqnarray}
where $\hat{H}_\alpha$ are intrastate vibrational Hamiltonians for the three electronic levels and $\hat{V}_\mathrm{int}$ is the light-matter dipole interaction Hamiltonian,
\begin{eqnarray}
 \hat{V}_\mathrm{int}  &= &  \sum_{a=1}^3 \hat{V}_{\mathrm{int},a}  \nonumber \\
  & = & - \sum_{a=1}^3 \lambda_a E_a(t-T_a)  \nonumber \\
  & & \times \left(\hat{\mu} \ee^{i \Omega_a (t - T_a) -  i\kk_a \rr} + \hat{\mu}^\dg \ee^{-i \Omega_a (t - T_a) + i \kk_a \rr} \right) \ , \nonumber \\ & & \label{eq: Vint expression}
\end{eqnarray}
where $\lambda_a$ are the electric field amplitudes, $E_a(t)$ the normalized pulse envelope functions, and $\Omega_a$ and $T_a$ the carrier frequencies and central times of the pulses.
 The carrier-envelope phases of the pulses are purposely set to zero in Eq. (\ref{eq: Vint expression}), for the reason that the relative phases of the pulses were not controlled in the experiments.\cite{SKFA05,SA11} Instead, the intensity ($\propto \lambda_1^2 \lambda_2^2 \lambda_3^2$) of the field-induced polarization was measured; for this homodyne-detected signal the inter-pulse optical phase-shifts are irrelevant and do not need to be actively stabilized.
The operator $\hat{V}_\mathrm{int}$ is written according to the rotating wave approximation, which implies that the transition operators $\hat{\mu}$ and $\hat{\mu}^\dg$, combined with the oscillatory terms $ \ee^{i \Omega_a (t - T_a)}$ and $ \ee^{-i \Omega_a (t - T_a)}$, describe electronic de-excitation and excitation, respectively,\cite{GED09,BVM12}
\begin{equation}
\hat{\mu} = \mu_{BX} |X\ket \bra B| + \mu_{EB} |B \ket \bra E| \ ,
\end{equation}
where $\mu_{BX}$ and $\mu_{EB}$ are the real-valued transition dipole moments (TDM) for the $B \longleftarrow X$ and $E \longleftarrow B$ electronic transitions, assumed to be dependent only on the I--I vibrational coordinate.

For weak fields, the wavefunction resulting from the interactions with the sequence of pulses can be expressed using third-order perturbation theory as\cite{mukamel}
\begin{equation}
|\Psi,t\ket = \sum_{n=0}^3 \left|\Psi^{(n)},t \right\ket \ ,  \label{eq: Pert Exp I}
\end{equation}
\begin{eqnarray}
\left| \Psi^{(n)},t \right\ket & = &  \left(-\frac{i}{\hbar}\right)^n \int_{-\infty}^t \dd\tau_n \cdots \int_{-\infty}^{\tau_2} \dd\tau_1  \nonumber \\
 & & \times \ee^{-\frac{i}{\hbar} \hat{H}_0 t}\hat{V}_\mathrm{int}(\tau_n) \cdots \hat{V}_\mathrm{int}(\tau_1) |\Psi,-\infty\ket \ ,\nonumber \\ & &  \label{eq: Pert Exp II}
\end{eqnarray}
where $\hat{V}_\mathrm{int}(\tau)$ is the time-dependent interaction representation of the operator $\hat{V}_\mathrm{int}$,
\begin{equation}
\hat{V}_\mathrm{int}(\tau) = \ee^{\frac{i}{\hbar} \hat{H}_0\tau} \hat{V}_\mathrm{int} \ee^{- \frac{i}{\hbar} \hat{H}_0\tau} \ ,
\end{equation}
and $|\Psi,-\infty\ket$ is the initial wavefunction, at infinite time before the first interaction with the light pulses, defined as the ground state of the field-free Hamiltonian $\hat{H}_0$, $|\Psi,-\infty\ket = |X\ket |\chi_0\ket $.  Clearly, $\left| \Psi^{(0)},t \right\ket = \ee^{-i \varepsilon_0^X t /\hbar} |\Psi,-\infty\ket$, where $\varepsilon_0^X$ is the energy of the vibrational ground state $|\chi_0\ket$. The higher order perturbative terms are found by replacing Eq. (\ref{eq: Vint expression}) into the expansion of Eqs. (\ref{eq: Pert Exp I}) and (\ref{eq: Pert Exp II}), and can be written compactly as
\begin{equation}
\left| \Psi^{(1)},t \right\ket = \sum_{a=1}^3 |B\ket \left| \chi_a^B,t \right\ket \ee^{i \kk_a \rr} \ ,  \label{eq: Psi1 from PT}
\end{equation}
\begin{eqnarray}
\left| \Psi^{(2)},t \right\ket & = & \sum_{a,b=1}^3 |X\ket \left| \chi_{ba}^{XB},t \right\ket \ee^{i(\kk_a - \kk_b)\rr}  \nonumber \\
 & + & \sum_{a,b=1}^3 |E\ket \left| \chi_{ba}^{EB},t \right\ket \ee^{i(\kk_a + \kk_b) \rr} \ ,
\end{eqnarray}
\begin{eqnarray}
\left| \Psi^{(3)},t \right\ket & = & \sum_{a,b,c=1}^3 |B\ket \left| \chi_{cba}^{BXB},t \right\ket \ee^{i(\kk_a-\kk_b+\kk_c)\rr} \nonumber \\
    &  +  &  \sum_{a,b,c=1}^3 |B\ket \left| \chi_{cba}^{BEB},t \right\ket \ee^{i(\kk_a + \kk_b - \kk_c)\rr} \ . \label{eq: Psi3 from PT}
\end{eqnarray}
The notation $\left| \chi_{...ba}^{...\beta \alpha}, t \right\ket$ is a shorthand to indicate the vibrational wave packet created by the transitions $X \longrightarrow \alpha \longrightarrow \beta \longrightarrow ...$ induced in sequence by the pulses $a$, $b$,... The third-order polarization induced by the interactions with the laser fields is 
\begin{eqnarray}
P^{(3)}(t) & = & \left\bra \Psi^{(3)},t \left| \hat{\mu} + \hat{\mu}^\dg \left| \Psi^{(0)},t \right\ket \right. \right. \nonumber \\
 & + & \left\bra \Psi^{(2)},t \left| \hat{\mu} +\hat{\mu}^\dg \left| \Psi^{(1)},t \right\ket \right. \right. \nonumber \\
 & + & \left\bra \Psi^{(1)},t \left| \hat{\mu} + \hat{\mu}^\dg \left| \Psi^{(2)},t \right\ket \right. \right. \nonumber \\
 & + & \left\bra \Psi^{(0)},t \left| \hat{\mu} + \hat{\mu}^\dg \left| \Psi^{(3)},t \right\ket \right. \right. 
 \label{eq: P3 from PT}
\end{eqnarray}
and propagates along several phase-matching directions,\cite{mukamel}
\begin{eqnarray}
 P^{(3)}(t) & = & P_\mathrm{S}^{(3)}(t) \ee^{i (-\kk_1 + \kk_2 + \kk_3)\rr} \nonumber \\
  & + & P_\mathrm{AS}^{(3)}(t) \ee^{i (+ \kk_1 - \kk_2 + \kk_3) \rr} \nonumber \\
  & + & P_\mathrm{other}^{(3)}(t) \ .
\end{eqnarray}
We focus on the components along the Stokes and anti-Stokes directions, $P_\mathrm{S}^{(3)}(t)$ and $P_\mathrm{AS}^{(3)}(t)$, and do not consider here the polarization $P_\mathrm{other}^{(3)}(t)$ propagating along all other directions; the components of interest are evaluated by replacing Eqs. (\ref{eq: Psi1 from PT})-(\ref{eq: Psi3 from PT}) into Eq. (\ref{eq: P3 from PT}) and keeping all terms proportional to $\ee^{i (-\kk_1 + \kk_2 + \kk_3)\rr}$ or $\ee^{i (+ \kk_1 - \kk_2 + \kk_3) \rr}$. In the experiments, the central times of the first two pulses (pump), $T_1$ and $T_2$, are fixed and the polarization is detected as function of the arrival time of the third (probe) pulse. We adopt the convention that $T_1=0$, so that the central time of the third pulse is regarded as a pump-probe delay, $T_\Pro = T_3$.

At this stage we make an approximation, based on the conditions of experimental setup of Refs. \onlinecite{SKFA05} and \onlinecite{SA11}, in which each pulse was tailored to specific electronic transitions: The carrier frequencies $\Omega_1$ and $\Omega_2$ were chosen to be resonant only with the $B \longleftrightarrow X$ transition, which implies that the TDM $\mu_{EB}$ can be set to zero in the operators $\hat{V}_{\mathrm{int},1}$ and $\hat{V}_{\mathrm{int},2}$ of Eq. (\ref{eq: Vint expression}); $\Omega_3$ was chosen to be resonant with the $E \longleftrightarrow B$ transition at the outer turning point on the $B$ surface, so that $\mu_{XB}$ is set to zero in the operator $\hat{V}_{\mathrm{int},3}$. Strictly speaking, the experimental value of $\Omega_3 = 20000\,\mathrm{cm}^{-1}$ is also in resonance with the high wavelength side of the $B \longleftarrow X$ absorption spectrum;\cite{KAMAP05,SA11} however, the absorption intensity is rather low and the Feynman pathways where the third pulse is associated with the $B \longleftrightarrow X$ transition vanish for non-overlapping pump and probe pulses. Moreover, Feynman pathways which involve $E \longleftrightarrow B$ transitions are favored by the fact that the TDM $\mu_{EB}$ is roughly twice as large as $\mu_{BX}$.\cite{T11,ABLPP11}
Under the just stated experimental conditions, the Stokes signal as a function of pump-probe delay $P_\mathrm{S}^{(3)}(t,T_\Pro)$ is mainly determined by a single Feynman pathway,\cite{SKFA05}
\begin{eqnarray}
 P_\mathrm{S}^{(3)}(t,T_\Pro) &  & \nonumber \\
&  & \hspace{-1.5cm}  = \left\bra \chi_1^B,t \left| \mu_{EB} \left| \chi_{32}^{EB},t \right\ket  \right. \right. \nonumber \\
& & \hspace{-1.5cm}  = -\frac{i}{\hbar} \int_{-\infty}^t \left\bra \chi_1^B,t \left| \mu_{EB} \ee^{-\frac{i}{\hbar} \hat{H}_E(t-\tau)} \mu_{EB} \right| \chi_2^B,\tau \right\ket  \nonumber \\
  & &   \times \lambda_3 E_3 \left(\tau - T_\Pro \right) \ee^{- i \Omega_3 (\tau - T_\Pro)} \dd\tau \ .  \label{eq: P3S formula}
\end{eqnarray}
The interpretation of Eq. (\ref{eq: P3S formula}) is straightforward: The first two pulses create the wave packets $\chi_1^B$ and $\chi_2^B$ on the $B$ state surface, 
\begin{eqnarray}
\left|\chi_a^B,t \right\ket & = & -\frac{i}{\hbar} \int_{-\infty}^t \ee^{-\frac{i}{\hbar} \hat{H}_B (t-\tau)} \mu_{BX} |\chi_0\ket \ee^{-\frac{i}{\hbar} \varepsilon_0^X \tau} \nonumber \\
 & & \times \lambda_a E_a(\tau - T_a) \ee^{-i\Omega_a (\tau - T_a)} \dd\tau \ , \  a = 1,2 \ , \nonumber \\  & & 
\end{eqnarray}
which are then probed as a function of $T_\Pro$ by the electronic excitation of $\chi_2^B$ to the state $E$, followed by spontaneous Raman scattering back to the $B$ surface, where the quantum mechanical overlap with $\chi_1^B$ determines the resulting measurable polarization. 

The corresponding expression for the anti-Stokes time resolved signal, $P_\mathrm{AS}^{(3)}(t,T_\Pro)$, is immediately obtained from Eq. (\ref{eq: P3S formula}) by swapping $\chi_1^B$ and $\chi_2^B$. The transitions involved in the CSRS and CARS processes are depicted in Figs. \ref{fig: Pulses}(a) and (b), respectively. According to Eq. (\ref{eq: P3S formula}), the calculation of the polarization would involve a number of quantum dynamical runs on the $E$ state potential energy surface. Such computations can be avoided by exploiting the analysis given below.

\subsection{Analysis of the signal and computational evaluation}
\label{sec: analysis}
In the resonance Raman process under experimental conditions, the propagation time on the $E$ state, $t - \tau$ in Eq. (\ref{eq: P3S formula}), is limited by the probe pulse duration and the electronic dephasing time $1 / \Gamma$, which is described phenomenologically as an exponential damping factor $\ee^{-\Gamma (t - \tau)}$. In the actual crystal, the electronic dephasing between the states $B$ and $E$ occurs on a time scale of the order of $\approx 100$\,fs\cite{SKFA05,SA11} and is driven by both the lattice vibrations and the nonadiabatic transitions to electronic states close to $E$.\cite{ZSA96,BC97} 

This situation allows one to avoid a full treatment of the $\mathrm{I_2:Kr}$ dynamics following the $E \longleftarrow B$ transition. As shown in paper I,\cite{PCB18A} the system vibration (i. e. the I--I stretch) has a large frequency compared to the remaining bath modes, whose vibrational periods are long compared to $1 / \Gamma$ ($\approx 700\,\mathrm{fs \ vs \ } \approx 100\,\mathrm{fs}$). On this basis, the bath can be approximated as being at rest during the evolution on the $E$ state, provided that the probe pulse duration is sufficiently short. In this case the time evolution operator for the transient sojourn in the $E$ state can be expressed in the basis $\{ \bar{\phi}_n \}$ of the eigenstates of $\mathrm{I_2}$ on the $E$ state,
\begin{eqnarray}
\hat{H}_\mathrm{I-I}^{(E)} \bar{\phi}_n(q) & = & \left(- \frac{\hbar \omega}{2} \frac{\partial^2}{\partial q^2} + V_\mathrm{I-I}^{(E)}(q) \right) \bar{\phi}_n(q) \nonumber \\ 
 & = & \bar{\varepsilon}_n \bar{\phi}_n(q) \ ,  \label{eq: chi_E}
\end{eqnarray}
as
\begin{equation}
 \ee^{- \frac{i}{\hbar} \hat{H}_E (t - \tau)} \approx \sum_n \left| \bar{\phi}_n \right\ket \ee^{-\frac{i}{\hbar} \left( \bar{\varepsilon}_n - i \hbar \Gamma \right) (t - \tau)} \left\bra \bar{\phi}_n \right| \ ,  \label{eq: propagator}
\end{equation}
where $q$ is the dimensionless I--I stretching mode and $\omega$ is its normal frequency (they are denoted $q_1$ and $\omega_1$ in paper I\cite{PCB18A}). The time evolution operator is defined by augmenting the iodine $E$ state Hamiltonian with the imaginary electronic dephasing rate $\hat{H}_\mathrm{I-I}^{(E)} \longrightarrow \hat{H}_\mathrm{I-I}^{(E)} - i \hbar \Gamma$. 

The Hamiltonian $\hat{H}_\mathrm{I-I}^{(E)}$ refers to the chromophore subsystem and is obtained from the full dimensional Hamiltonian by setting the bath normal coordinates $q_j$ to the equilibrium geometry of the ground state $X$ ($q_j = 0$). The analogous $\mathrm{I_2}$ Hamiltonian for the $B$ state (see Eq. (27) of  paper I\cite{PCB18A}) defines the vibrational energy levels of the chromophore via the Schr\"{o}dinger equation
\begin{eqnarray}
\hat{H}_\mathrm{I-I}^{(B)} \phi_j(q) & = & \left(- \frac{\hbar \omega}{2} \frac{\partial^2}{\partial q^2} + V_\mathrm{I-I}^{(B)}(q) \right) \phi_j(q) \nonumber \\ 
 & = & \varepsilon_j \phi_j(q) \ . \label{eq: chi_B}
\end{eqnarray}

The wave packets $\chi_1^B$ and $\chi_2^B$ are expanded as a sum of products of $\mathrm{I_2}$ vibrational wavefunctions of the $B$ state and associated single-hole functions,
\begin{equation}
\chi_a^B(q,\qq_\mathrm{bath},t) = \sum_j \ee^{- \frac{i}{\hbar} \varepsilon_j t} \phi_{j}(q) \psi_{aj}(\qq_\mathrm{bath},t) \ ,   \label{eq: psi expand}
\end{equation}
where $a=1,2$ and $\qq_\mathrm{bath}$ includes all bath coordinates. Replacing Eqs. (\ref{eq: propagator}) and (\ref{eq: psi expand}) into Eq. (\ref{eq: P3S formula}) one obtains 
\begin{eqnarray}
 P_\mathrm{S}^{(3)}(t,T_\Pro) & = & \nonumber \\
 & & \hspace{-1.5cm}  - \frac{i}{\hbar} \sum_{j,l} \sum_n \int_{-\infty}^t \mathrm{d}\tau \lambda_3 E_3(\tau - T_\Pro) \ee^{-i\Omega_a(\tau - T_\Pro)} \nonumber \\
 & & \hspace{-1.5cm} \times \bra \phi_j |\mu_{EB} | \bar{\phi}_n \ket  \bra \bar{\phi}_n| \mu_{EB} | \phi_l \ket  \ee^{-\frac{i}{\hbar} \left(\bar{\varepsilon}_n - i\hbar \Gamma \right) (t - \tau)} \nonumber \\ 
 & &\hspace{-1.5cm}  \times C_{jl}^{(12)}(t,\tau)  \ee^{\frac{i}{\hbar} \varepsilon_j t} \ee^{-\frac{i}{\hbar} \varepsilon_l \tau}   \label{eq: P3 expand}
\end{eqnarray}
where $C_{jl}^{(12)}(t,\tau) = \bra \psi_{1j}(t) | \psi_{2l}(\tau) \ket$ is the non-oscillatory cross-correlation matrix element in the basis of the vibrational levels of the $B$ state. 
Eq. (\ref{eq: P3 expand}) is used to compute the time-resolved Raman spectra of Sect. \ref{sec: calculated_signal}. The evaluation of the third-order polarization does not require any high-dimensional wave packet propagation on the $E$ state surface. As explained in Sect. \ref{sec: comp. det.}, quantum dynamical calculations are used only to obtain the wavefunctions $\chi_1^B$ and $\chi_2^B$ of Eq. (\ref{eq: psi expand}), which evolve on the $B$ state surface and are necessary for the computation of the matrix elements $C_{jl}^{(12)}(t,\tau)$.

In order to derive an expression for the interpretation of the third-order signals, a further approximation of the cross-correlation matrix is considered.
Note that the phase factors $\ee^{-\frac{i}{\hbar} \varepsilon_j t}$ associated with the bath wavefunctions have been made explicit in Eq. (\ref{eq: psi expand}), so that the overlaps $C_{jl}^{(12)}(t,\tau)$ between bath wavefunctions are expected to vary slowly as a function of $t - \tau$, which is the dephasing-limited propagation time on the $E$ state surface. In this way the polarization can be calculated by approximating the cross-correlation matrix with the wave packet coherence matrix in the energy representation, evaluated at the midpoint between $t$ and $\tau$, 
\begin{equation}
C_{jl}^{(12)}(t,\tau) \approx C_{jl}^{(12)}\left(\frac{t + \tau}{2},\frac{t + \tau}{2}\right) \equiv C_{jl}^{(12)}\left( \frac{t + \tau}{2} \right) \ .  \label{eq: C approx}
\end{equation}
This approximation has two advantages: first, it allows one to avoid the computation of the two-time cross-correlation matrix and the storage in memory of the full wavefunction; second, it facilitates the interpretation of the spectrum by providing an expression of the spectroscopic signal in which the coherence matrix $C_{jl}^{(12)}$ appears explicitly. The time-resolved CSRS spectrum is obtained by evaluating the Fourier transform of $P_\mathrm{S}^{(3)}(t,T_\Pro)$ at the Raman scattering frequency $\OmegaR$,
\begin{equation}
 \widetilde{P}_\mathrm{S}^{(3)}(\OmegaR,T_\Pro) \sim \int_{-\infty}^{+\infty} P_\mathrm{S}^{(3)}(t,T_\Pro) \ee^{i \OmegaR t} \mathrm{d}t \ . \label{eq: P FT}
\end{equation}
Substituting Eqs. (\ref{eq: P3 expand}) and (\ref{eq: C approx}) into Eq. (\ref{eq: P FT}), and changing the integration variable as $t \longrightarrow 2t - \tau$ one obtains 
\begin{eqnarray}
\hspace{-1cm} \widetilde{P}_\mathrm{S}^{(3)}(\OmegaR,T_\Pro) & = & \nonumber \\
 & & \hspace{-2cm} -\frac{2i \lambda_3}{\hbar} \sum_{jl}  \int_{-\infty}^{+\infty} \mathrm{d}t \exp \left[2 i \left(\Omega_R - \frac{\bar{\varepsilon}_n - \varepsilon_j}{\hbar} + i\hbar \Gamma \right) t \right]   \nonumber  \\
 & & \hspace{-2cm} \times  C_{jl}^{(12)}(t) \sum_n \bra \phi_j |\mu_{EB} | \bar{\phi}_n \ket  \bra \bar{\phi}_n| \mu_{EB} | \phi_l \ket \nonumber \\
 & & \hspace{-2cm} \times \int_{-\infty}^t \mathrm{d}\tau E_3(\tau - T_\Pro) \ee^{i \Omega_3 T_\Pro} \nonumber \\
 & & \hspace{-2cm} \times \exp\left[i \left( \frac{\bar{\varepsilon}_n - \varepsilon_j}{\hbar} - \OmegaR + \frac{\bar{\varepsilon}_n - \varepsilon_l}{\hbar} - \Omega_3 - 2i\hbar \Gamma \right) \tau \right] \ . \nonumber \\
 & &  \label{eq: Spec2D_1}
\end{eqnarray}
Eq. (\ref{eq: Spec2D_1}) might appear cumbersome but it is a readily usable working equation: As explained in Sect. \ref{sec: comp. det.}, the matrix $C_{jl}^{(12)}(t)$ is constructed from the density matrices obtained in four quantum dynamical runs, without the necessity of storing the full time-dependent wavefunctions; the iodine energy levels and the Franck-Condon factors $\bra \phi_j |\mu_{EB} | \bar{\phi}_n \ket  \bra \bar{\phi}_n| \mu_{EB} | \phi_l \ket$ are obtained by one-dimensional Hamiltonian diagonalizations. Moreover, the resonance conditions -- by which the rapidly oscillating terms in the time integrals of Eq. (\ref{eq: Spec2D_1}) are removed -- can be readily inferred as $\hbar \OmegaR \approx \bar{\varepsilon}_n - \varepsilon_j$, $\hbar \Omega_3 \approx \bar{\varepsilon}_n - \varepsilon_l$. 

In more physical terms, in the signal of Eq. (\ref{eq: Spec2D_1}) the vibrational coherence between the energy levels $\phi_l$ and $\phi_j$ is monitored via absorption to the $E$ state at the probe frequency and subsequent spontaneous scattering to the $B$ state. The Raman shift $\omega_\mathrm{S} = \Omega_3 - \OmegaR$ is positive if the arrival state has a higher energy than the probed state, $\varepsilon_j > \varepsilon_l$; conversely, $\omega_\mathrm{S} < 0$ if $\varepsilon_j < \varepsilon_l$. The elements of the coherence matrix $C_{jl}^{(12)}(t)$ which contribute to the signal for a given (positive or negative) Raman shift are the ones displaced by $\pm \hbar \omega_\mathrm{S}$ from the energy diagonal. 
In the Appendix, the time- and frequency-resolved signals calculated using Eq. (\ref{eq: Spec2D_1}) -- or, equivalently, Eq. (\ref{eq: Spec2D_4}) reported below -- are compared with the spectra obtained from the more accurate Eq. (\ref{eq: P3 expand}).

Further insight is obtained by re-writing Eq. (\ref{eq: Spec2D_1}) more compactly. Defining $\widetilde{C}_{jl}^{(12)}(t) = C_{jl}^{(12)}(t) \ee^{i \left( \varepsilon_j - \varepsilon_l \right) t / \hbar} = \left\bra \chi_1^B,t \left| \phi_j \left\ket \left\bra \phi_l \left| \chi_2^B,t \right\ket \right. \right. \right. \right.$, $S_{n,jl} = \bra \phi_j |\mu_{EB} | \bar{\phi}_n \ket  \bra \bar{\phi}_n| \mu_{EB} | \phi_l \ket$, and $\hbar \Delta_{n,jl} = \left(2 \bar{\varepsilon}_n - \varepsilon_j - \varepsilon_l \right)$ one gets

\begin{eqnarray}
 \widetilde{P}_\mathrm{S}^{(3)}(\OmegaR,T_\Pro) & = &  -\frac{2i \lambda_3}{\hbar}  \ee^{i \Omega_3 T_\Pro}  \nonumber \\
  & & \hspace{-2cm} \times \sum_{jl}  \int_{-\infty}^{+\infty} \mathrm{d}t \widetilde{C}_{jl}^{(12)}(t)  \sum_n S_{n,jl}  \ee^{i (2 \OmegaR - \Delta_{n,jl} + 2i\hbar\Gamma) t} \nonumber \\ 
  & & \hspace{-2cm} \times \int_{-\infty}^t \mathrm{d}\tau E_3(\tau - T_\Pro) \ee^{i (\Delta_{n,jl} - \OmegaR - \Omega_3 - 2i\hbar\Gamma) \tau} \ , \nonumber \\
  & &  \label{eq: Spec2D_2}
\end{eqnarray}
and, upon the variable change $\tau \longrightarrow \tau + T_\Pro$,
\begin{eqnarray}
\widetilde{P}_\mathrm{S}^{(3)}(\OmegaR,T_\Pro)  & = &  -\frac{2i \lambda_3}{\hbar} \ee^{i \OmegaR T_\Pro} \sum_{jl} \int_{-\infty}^{+\infty} \mathrm{d}t \widetilde{C}_{jl}^{(12)}(t)  \nonumber \\
 & & \hspace{-2cm} \times  \sum_n S_{n,jl}  \ee^{i (2 \OmegaR - \Delta_{n,jl} + 2i\hbar\Gamma) (t - T_\Pro)} \nonumber \\
 & & \hspace{-2cm} \times \int_{-\infty}^{t - T_\Pro} \mathrm{d}\tau E_3(\tau) \ee^{i (\Delta_{n,jl} - \OmegaR - \Omega_3 - 2i\hbar\Gamma) \tau} \ . \label{eq: Spec2D_3}
\end{eqnarray}
The expression for the time-dependent Raman signal can be finally arranged in the form of a convolution,
\begin{eqnarray}
\widetilde{P}_\mathrm{S}^{(3)}(\OmegaR,T_\Pro) & = & -\frac{2i\lambda_3}{\hbar} \ee^{i \OmegaR T_\Pro} \sum_{jl} \widetilde{C}_{jl}^{(12)} \star W_{lj}^{\Omega_3, \OmegaR}(T_\Pro) \nonumber \\
 & = &  -\frac{2i\lambda_3}{\hbar} \ee^{i \OmegaR T_\Pro} \mathrm{Tr}\left[ \widetilde{\mathbf{C}}^{(12)} \star \mathbf{W}^{\Omega_3, \OmegaR}(T_\Pro) \right] \ ,  \nonumber \\
 & & \label{eq: Spec2D_4}
\end{eqnarray}
where
\begin{equation}
 \widetilde{C}_{jl}^{(12)} \star W_{lj}^{\Omega_3, \OmegaR}(T_\Pro) = \int_{-\infty}^{+\infty} \mathrm{d}t \widetilde{C}_{jl}^{(12)}(t) W_{lj}^{\Omega_3, \OmegaR}(T_\Pro - t) \ .
\end{equation}
In Eq. (\ref{eq: Spec2D_4}) $\widetilde{C}_{jl}^{(12)}(t)$ is the time-dependent coherence between the wave packets $\chi_1^B$ and $\chi_2^B$ in the energy representation, and contains all dynamical information about the motion on the $B$ state. $W_{jl}^{\Omega_3, \OmegaR}(t)$ are \lq time and frequency gate' (TFG) functions which do not depend on the evolving wave packets (i. e. on the pump stage), but only on the system Hamiltonian and the probe field, and describe the intrinsic probability of going from the level $\phi_l$ to the level $\phi_j$ via the probe-scattering sequence. Its explicit expression is given as
\begin{eqnarray}
 W_{lj}^{\Omega_3, \OmegaR}(t) & = & \sum_n S_{n,jl} \ee^{i \left( \Delta_{n,jl} - 2\OmegaR  -2 i\hbar \Gamma \right) t} \nonumber \\ 
  & & \hspace{-1.5cm} \times \int_t^{+\infty} E_3(-\tau) \ee^{- i \left( \Delta_{n,jl} - \OmegaR - \Omega_3 - 2 i\hbar \Gamma \right) \tau} \mathrm{d}\tau \ ,
\end{eqnarray}
and the integral over $\tau$ can be evaluated analytically for a number of pulse shapes. 

Eq. (\ref{eq: Spec2D_4}) is akin to the expressions for the time-resolved fluorescence, two- and three-pulse signals obtained under the doorway-window formalism.\cite{UC97,MCH97,CMM00,GPD02,GED04,GK08} In this sense, $ \widetilde{C}_{jl}(t)$ is a \lq doorway' coherence matrix, created by the pump pulse pair, and $W_{jl}^{\Omega_3, \OmegaR}(t)$ is a \lq window' function which also incorporates the approximate short-time dynamics on the $E$ state. The quantities $W_{lj}^{\Omega_3, \OmegaR}(t)$ decay rapidly as a function of time and, according to Eq. (\ref{eq: Spec2D_4}), provide a \lq time-gate' for the transient observation of the evolving vibrational coherence: The signal $\widetilde{P}_\mathrm{S}^{(3)}(\OmegaR,T_\Pro)$ is a detector of the coherence between $\chi_1^B$ and $\chi_2^B$ at time $T_\Pro$. The region of the coherence matrix that is effectively probed is defined by both the Raman shift, which selects a specific energy diagonal, and the TFG function $W_{lj}^{\Omega_3, \OmegaR}(t)$, which restricts the selection to the states with large Franck-Condon factors $S_{n,jl}$ and in resonance with the virtual states via the probe and Raman transitions.

As a final remark, we emphasize that the expression for the signal, Eq. (\ref{eq: Spec2D_4}), contains the trace of a product of two matrices, and as such it is invariant with respect to changes of representations. As shown in Paper I,\cite{PCB18A} different representations, like the phase space picture or the $(q,q^\prime)$ basis, provide a multifaceted view of the wave packet dynamics. For example, pump-probe signals, which can be described by expressions similar to Eq. (\ref{eq: Spec2D_4}), are often conveniently interpreted using the phase space representation.\cite{LZAM95,LFM96,BC97,SC99,ZV14} In the present four-wave-mixing signals, however, the probe coherence matrix $\widetilde{\mathbf{C}}$ is non-Hermitian, therefore its Wigner transform is complex and not straightforward to analyze; therefore the energy representation used in the derivation seems the most appropriate for this application.

\section{Computational Details}
\label{sec: comp. det.}

\begin{figure}[b!]
\centering
\includegraphics[scale=0.15]{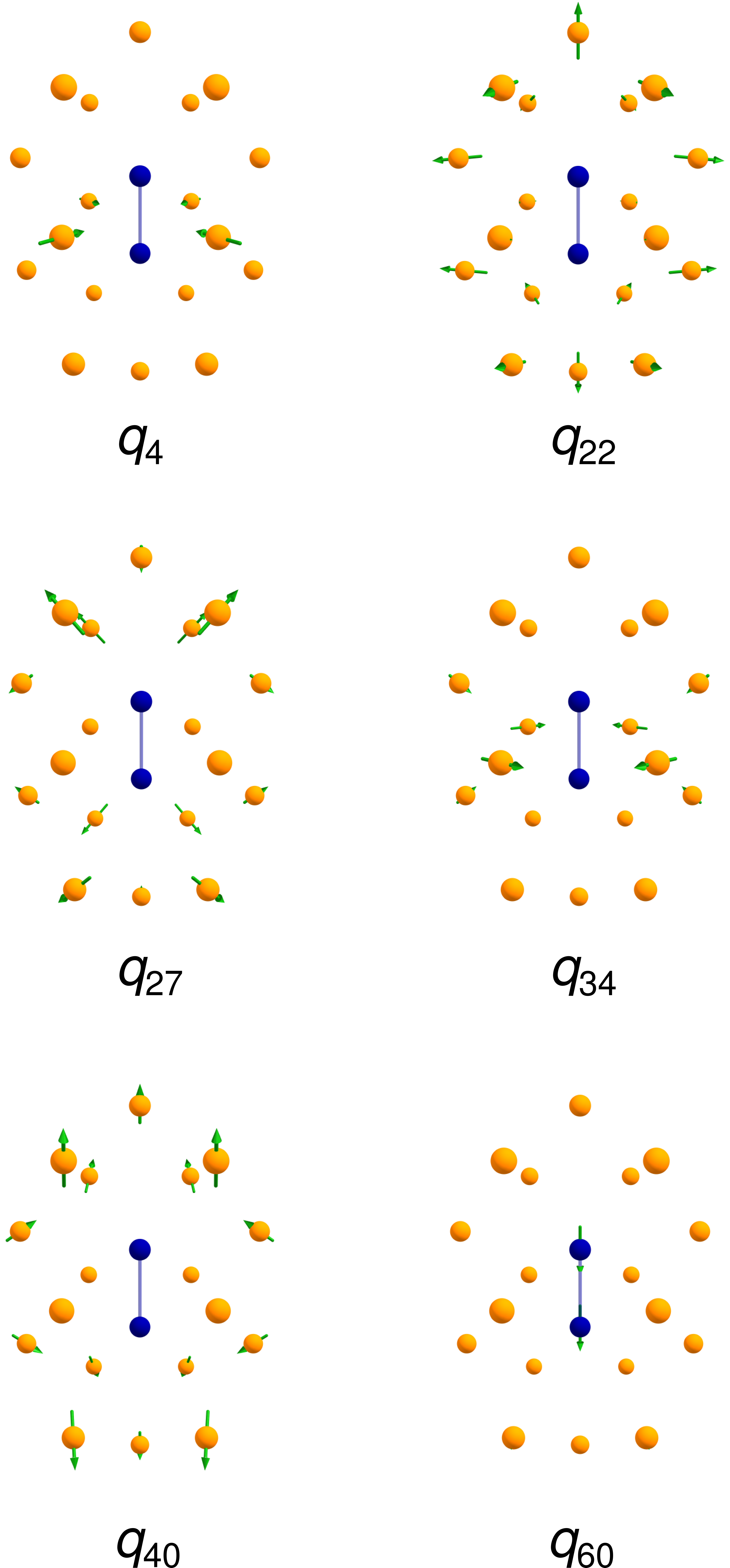}
\caption{The normal modes of the $\mathrm{I_2 Kr_{18}}$ cluster included in quantum dynamical calculations, in addition to the I--I stretching mode $q$. } 
\label{fig: modes}
\end{figure}

Quantum dynamical calculations were performed for the $\mathrm{I_2 Kr_{18}}$ cluster using the seven-dimensional Hamiltonian described in paper I,\cite{PCB18A} which includes the I--I stretching mode ($q$) and the six bath modes depicted in Fig. \ref{fig: modes}. The complete list of parameters defining the $X$, $B$ and $E$ state Hamiltonians is reported in Paper I. As described in the companion work,\cite{PCB18A} the potentials were derived from force fields consisting of atom-atom pairwise interactions; for the $E$ state, the I--I interaction potential was taken from Ref. \onlinecite{BC97} and the $E$ state potential minimum was adjusted to be $\mathrm{37000\,cm^{-1}}$ above the $X$ state minimum.

The wave packets $\chi_1^{B}$ and $\chi_2^{B}$ evolving on the B state surface were described using G-MCTDH wavefunctions according to the settings III of paper I (details are reported in Table \ref{tab: gmctdh}). The G-MCTDH equations of motion were integrated using a sixth-order Adam-Bashfort-Moulton integrator with variable step size and an integration accuracy $\varepsilon_\mathrm{int}=10^{-6}$. With these settings and propagation runs of 3.5\,ps the energy conservation, in absence of external fields, is obtained within the range $\mathrm{\pm 2\,cm^{-1}}$, which is fully appropriate for the spectral calculations of this work.  The wave packets are the components on the $B$ state of the first order solutions of the time-dependent Schr\"{o}dinger equation
\begin{equation}
 i\hbar \frac{\partial}{\partial t} \left|\chi_a^{B},t \right\ket = \left(\hat{H}_0 + \hat{V}_{\mathrm{int},a}  \right)  \left|\chi_a^B,t\right\ket \ , a = 1,2;
   \label{eq: TDSE psi alpha}
\end{equation}
In the actual calculations they were obtained by solving Eq. (\ref{eq: TDSE psi alpha}) using a weak interaction Hamiltonian of the form of Eq. (\ref{eq: Vint expression}), so that the total population transferred to the $B$ state by the two pump pulses was of the order of 0.02.  The initial state prior to the photoexcitation was chosen as the ground vibrational state of the $X$ state surface in the harmonic approximation.

\begin{table}[b!]
\begin{ruledtabular}
\begin{tabular}{cccc}
 Particle & Type & $N$ & $n$  \\ \hline
 $q_1$      &  DVR &  351  & 18  \\
 $(q_4,q_{22},q_{40})$ & GWP & -- & 33  \\
 $(q_{27},q_{34},q_{60})$ & GWP & -- &  19    
\end{tabular}
\end{ruledtabular}
\caption{Computational details of the G-MCTDH calculations. $N$ and $n$ are respectively the number of primitive DVR grid points and the number of SPFs used for each particle. The type GWP indicates the modes for which a Gaussian representation is used.}
\label{tab: gmctdh}
\end{table}

The time-resolved CARS and CSRS spectra were calculated using Eq. (\ref{eq: P3 expand}) for the third-order polarization. The two-time cross-correlation matrix elements $C^{(12)}_{jl}(t,\tau)$ can be evaluated from the G-MCTDH wavefunctions as
\begin{equation}
C^{(12)}_{jl}(t,\tau) = \left\bra \chi_1^B,t \left| \phi_j \left\ket \left\bra \phi_l \left| \chi_2^B,\tau \right\ket \right. \right. \right. \right. \mathrm{e}^{-\frac{i}{\hbar}\varepsilon_j t} \mathrm{e}^{\frac{i}{\hbar} \varepsilon_l \tau} \ . \label{eq: C12 eval}
\end{equation}
Note that the calculation of $C^{(12)}_{jl}(t,\tau)$ requires the storage in memory of the wavefunctions $\chi_1^B$ and $\chi_2^B$ on a dense time grid. This storage can be avoided if the more approximate Eq. (\ref{eq: Spec2D_4}) is used.

The TFG function matrix $W_{jl}^{\Omega_3, \OmegaR}(t)$, necessary to calculate the time- and frequency-dependent spectra according to Eq. (\ref{eq: Spec2D_4}), was calculated using the iodine eigenfunctions $\phi_j(q)$ and $\bar{\phi}_n(q)$ obtained by solving the Schr\"{o}dinger equations \ref{eq: chi_B} and \ref{eq: chi_E} on the same discrete variable representation (DVR) grid used for the wave packet propagation. 
The transition dipole moment $\mu_{EB}$ was taken as geometry-independent, a truncated cosine-squared was used as envelope function of the probe, 
\begin{equation}
 f_\Pro(t) = \left\{ \begin{array}{lc}
 0  & \mbox{ for } |t| > \Delta_t \\
 \cos^2\left(\frac{\pi t}{2 \Delta_t}\right) & \mbox{ for } |t| \le \Delta_t
\end{array}  \right. \ ,
\end{equation}
with $\Delta_t = 40\,\mathrm{fs}$, and the electronic dephasing rate was taken from experimental estimates as $\Gamma = (100\,\mathrm{fs})^{-1}$.\cite{SKFA05}

\section{Time-resolved coherent Raman signal for Schr\"{o}dinger cat states}
\label{sec: calculated_signal}

The interaction with the pair of laser pump pulses creates the perfectly coherent Schr\"{o}dinger cat superposition of Eq. (\ref{eq: psi_alpha}). The specifics of the preparation pulses may affect the way the iodine chromophore interacts with its surroundings, and the rate of decoherence between the wave packets $\chi_1^B$ and $\chi_2^B$. Even the mechanistic events which drive the transition to classicality may be different for different preparations. 

Below, we study the decoherence dynamics and its spectroscopic signatures for two different wave packet superpositions, \textbf{A} and \textbf{B}, prepared by two different pump sequences. The envelope function of Eq. (\ref{eq: Vint expression}) is taken as Gaussian,
\begin{equation}
E_a(t) = \exp\left(- \frac{t^2}{2\Delta_a^2}\right) \ ,  \label{eq: field Gaussian}
\end{equation}
and the products $\lambda_a \mu_{BX}$ are fixed to $10^{-4}$ hartrees. The field parameters for the cases \textbf{A} and \textbf{B} are in line with the pulses used in the experiments of Ref. \onlinecite{SA11} and are reported in Table \ref{tab: cat parameters}. The central frequency of the probe pulse is set to $20000\,\mathrm{cm}^{-1}$ as in most of the experimental measurements.

\begin{table}[h!]
\begin{ruledtabular}
\begin{tabular}{ccccc}
\shortstack{wave packet\\superposition} & $a$ & $T_a$ [fs] & $\Omega_a \ [\mathrm{cm}^{-1}]$ & $\Delta_a$ [fs] \\  \hline
\textbf{A}  &  1 &   0        & 17914 & 48  \\
                     &  2 & $-30$ & 17589 & 53  \\
            &  &  &  & \\
\textbf{B}  &  1 &   0        & 18251 & 18  \\
                     &  2 &   0        & 17589 & 16
\end{tabular}
\end{ruledtabular}
\caption{Parameters of the light pulses used to create the initial wave packets $\chi_a^B$ of the Schr\"{o}dinger cat superpositions \textbf{A} and \textbf{B} of Eq. (\ref{eq: psi_alpha}), for which the time-resolved coherent Raman spectra are calculated. The interaction Hamiltonian is defined according to Eqs. (\ref{eq: Vint expression}) and (\ref{eq: field Gaussian}). }
\label{tab: cat parameters}
\end{table}

\subsection{Superposition \textbf{A}} 
\label{sec: case A}

\begin{figure*}
\centering
\includegraphics[width=1.0\textwidth]{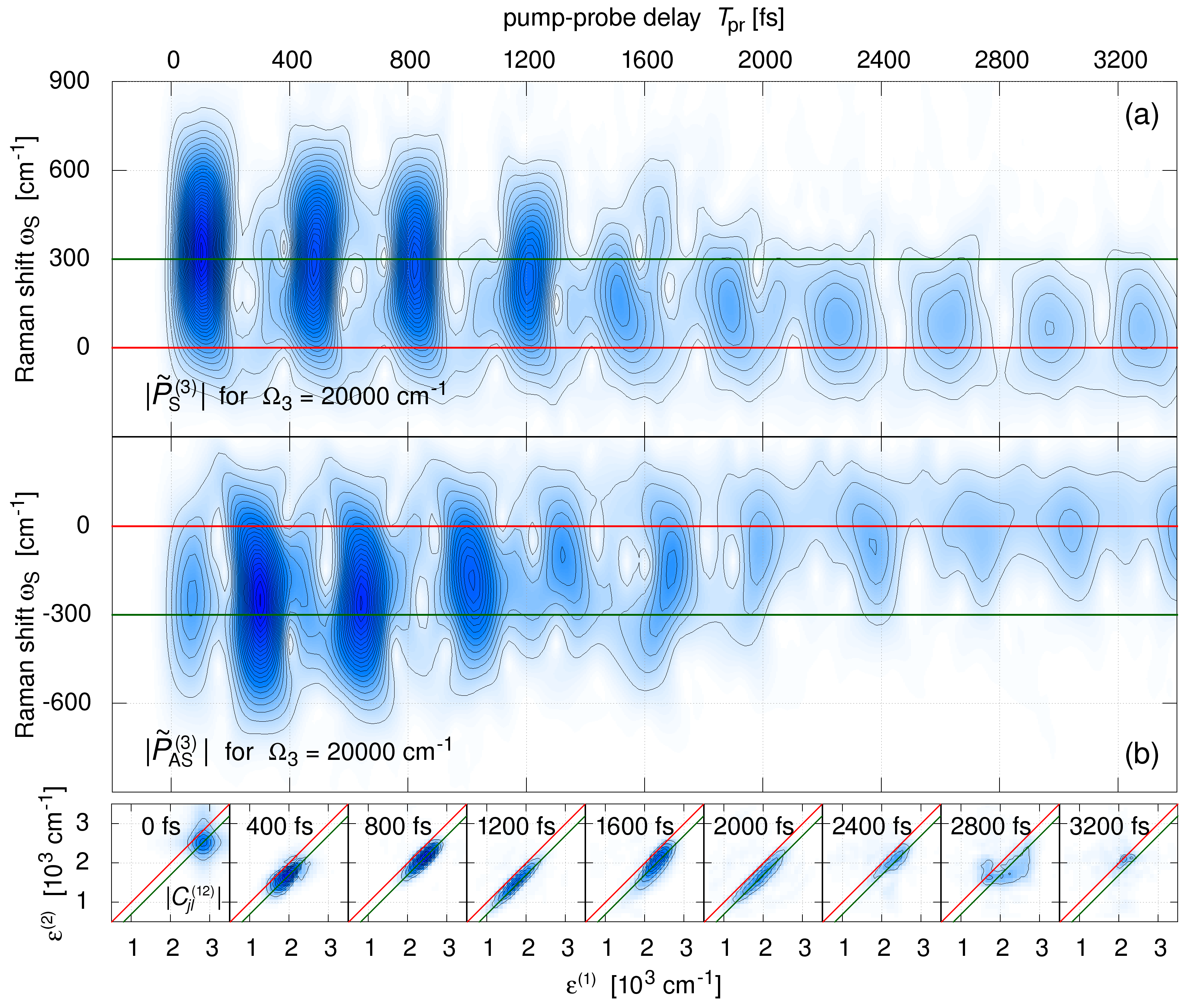}
\caption{The time-resolved coherent Stokes (a) and anti-Stokes (b) Raman spectra of the $\mathrm{I_2 Kr_{18}}$ cluster resulting from the initial wave packet superposition \textbf{A} (see Table \ref{tab: cat parameters}), as a function of probe delay and Raman shift $\omega_\mathrm{S} =  \Omega_3 - \OmegaR$, calculated using G-MCTDH quantum dynamics and for the probe frequency $ \Omega_3 = 20000\,\mathrm{cm}^{-1}$.   The bottom panels show contour maps of the absolute value of the reduced coherence matrix $|C_{jl}^{(12)}|$ as a function of time in the basis of the eigenstates of the embedded $\mathrm{I_2}$ chromophore in the electronic state $B$; the vibrational energies  $\varepsilon^{(1)}$ and $\varepsilon^{(2)}$ are relative to the ground vibrational energy on the $B$ state surface.} 
\label{fig: spectrum BB}
\end{figure*}

In this case the pair of pulses are rather narrow in energy [full width at half maximum $\mathrm{FWHM}\approx  180\,\mathrm{cm^{-1}}$ see Fig. \ref{fig: Pulses}(c)], and prepare two wave packets which are separated by $\approx 325\,\mathrm{cm^{-1}}$ in the energy scale. The resulting time-resolved CSRS and CARS spectra are shown in Fig. \ref{fig: spectrum BB}(a) and (b).
The spectra feature a sequence of Raman bands along the time axis, with the  spacing between subsequent peaks of $\approx 380\,\mathrm{fs}$ which corresponds to the vibrational frequency of $\approx 88\,\mathrm{cm}^{-1}$ associated with the vibrational quantum numbers $8 < \nu < 20$ for I--I stretch in the $B$ state. The first bands are broader, more intense and shifted to positive (Stokes) and negative (anti-Stokes) Raman shifts $\omega_\mathrm{S}$. Upon increase of the pump-probe delay, the bands lose intensity, they become slightly narrower and slowly converge towards the degenerate scattering region $\omega_\mathrm{S} \approx 0$.

 In the CSRS signal the first four bands ($T_\Pro  1.4\,\mathrm{ps}$) have an emission maximum at Raman shifts $\omega_\mathrm{S} \approx 300\,\mathrm{cm}^{-1}$ and their spectral width (FWHM) decreases monotonically from $470\,\mathrm{cm}^{-1}$ to $425\,\mathrm{cm}^{-1}$; at the same time the peak intensity declines to roughly half of the maximum of the first band. After 1.4\,ps the intensity drops rapidly by another factor two, the fifth and sixth bands of the sequence shift towards $\omega_\mathrm{S} \approx 200\,\mathrm{cm}^{-1}$ and their FWHM lowers to $\approx 410\,\mathrm{cm}^{-1}$.
 The two bands in the range $2.0 \,\mathrm{ps} < T_\Pro < 2.8\,\mathrm{ps}$ have a similar width but are significantly less intense and centered around a lower Raman shift $\omega_\mathrm{S} = 150\,\mathrm{cm}^{-1}$. The bands at long times peak at $\omega_\mathrm{S} < 70\,\mathrm{cm}^{-1}$ and have $\mathrm{FWHM}\approx 350\,\mathrm{cm}^{-1}$.

 In the CARS spectrum the earliest feature is a low intensity band which peaks at a delay time of $T_\mathrm{pr} = 60\,\mathrm{fs}$, i. e. when the action of the pump pulses is still not completely ceased. This signal is therefore a probe of the last step of creation of the Schr\"{o}dinger cat state. The sequence of bands which monitor the subsequent evolution of the wave packet superposition starts at $T_\mathrm{pr} \approx 200\,\mathrm{fs} $.

In the CARS spectrum, the band intensity decays slightly in the first 1.2\,ps (first three bands) and drops significantly after this time. The Raman shift at the first emission maximum is $\omega_S = -280\,\mathrm{cm}^{-1}$, that corresponds approximately to the negative of the first peak shift in the Stokes band sequence; at later times the bands gradually move towards $\omega_\mathrm{S} \approx 0$ and nearly localize after 2\,ps. The emission band width decreases from $430\,\mathrm{cm}^{-1}$ (for the first band) to $330-350\,\mathrm{cm}^{-1}$ (for the bands at $T_\Pro > 2\,\mathrm{ps}$).

As explained in Sect. \ref{sec: analysis}, the CSRS and CARS signals monitor the evolution of $C^{(12)}_{jl}(t)$ and $C^{(21)}_{jl}(t)$, which are simply each the Hermitian conjugate of the other, therefore they can be viewed as two different probes of the same quantum mechanical quantity. The coherence at different selected times is shown as a contour map in the bottom panels of Fig. \ref{fig: spectrum BB}, as a function of the $B$ state iodine energy levels $\varepsilon^{(1)}$ and $\varepsilon^{(2)}$ for the wave packets $\chi_1^B$ and $\chi_2^B$. In order to facilitate the readout of the coherence maps, the energy diagonal $\varepsilon^{(1)} = \varepsilon^{(2)}$ is shown in red in the plots, and the parallel line $\varepsilon^{(2)} = \varepsilon^{(1)} - 300\,\mathrm{cm}^{-1}$ is shown in green. The corresponding lines are traced in the spectra of Figs. \ref{fig: spectrum BB}(a) and (b). In fact, due to the resonance conditions implied by Eq. (\ref{eq: Spec2D_1}), the Stokes and anti-Stokes signals at Raman shifts $\pm \omega_\mathrm{S}$ are mostly contributed by the coherence along the line $\varepsilon^{(1)} - \varepsilon^{(2)} = \omega_\mathrm{S}$.
For example, since the central frequency of the preparation pulses differ by $350\,\mathrm{cm}^{-1}$, the maximum of the initial $\mathbf{C}$ matrix is displaced from the energy diagonal by the same amount, which also matches (in absolute value) the Raman shift of the first Stokes and anti-Stokes emission bands. 

The initial widths of $|C_{jl}(t=0)|$ are $535\,\mathrm{cm}^{-1}$ and $490\,\mathrm{cm}^{-1}$ along the $\varepsilon^{(1)}$ and $\varepsilon^{(2)}$ axes, respectively. These values are slightly larger than the width of the earliest Raman band in the CARS spectrum, as a consequence of the frequency filtering operated by the TFG function [see Eq. (\ref{eq: Spec2D_4})]. In contrast, the first CSRS bands have the same width as the coherence, suggesting that the Stokes phase-matching direction allows the detection of the coherence over a larger energy range. The non-equivalence between the two spectra is easily explained by Eq. (\ref{eq: Spec2D_4}) for the S signal: The AS spectrum is simply obtained by replacing the matrix $\mathbf{C}^{(12)}$ with its Hermitian conjugate and leaving the complex TFG matrix unchanged; since this matrix is symmetric but not Hermitian, no simple symmetry relations can be established between the CSRS and CARS spectra.

In the first 400\,fs the coherence shifts towards energies lower by $800-900\,\mathrm{cm}^{-1}$, due to the dissipation induced by the first molecule-cage collision. However, as already discussed in paper I, the dissipation in the first 2\,ps is non-monotonic, as shown by the motion the $\mathbf{C}$ matrix in Fig. \ref{fig: spectrum BB}, which oscillates and broadens along the energy diagonal.  The fact that the coherence is retained for the first four vibrational periods has a spectral counterpart in the higher intensity of the first four Stokes bands. An abrupt decoherence is observed in the coherence maps after 2\,ps, when the $\mathbf{C}$ matrix noticeably decays, shrinks and finally localizes on the energy diagonal; the same features are retrieved by the latest Stokes bands which are weak, narrow and gradually shift towards $\omega_\mathrm{S} = 0$. In the anti-Stokes spectrum, the dropoff of the signal observed around 1.2\,ps seems to be related to some dissipative bath motion which drives the coherence matrix towards low energies, more than $1000\,\mathrm{cm}^{-1}$ away from the initial maximum at time $t = 0\,\mathrm{fs}$ (see the coherence map at 1200\,fs in Fig. \ref{fig: spectrum BB}.

\begin{figure}[b!]
\centering
\includegraphics[scale = 0.3]{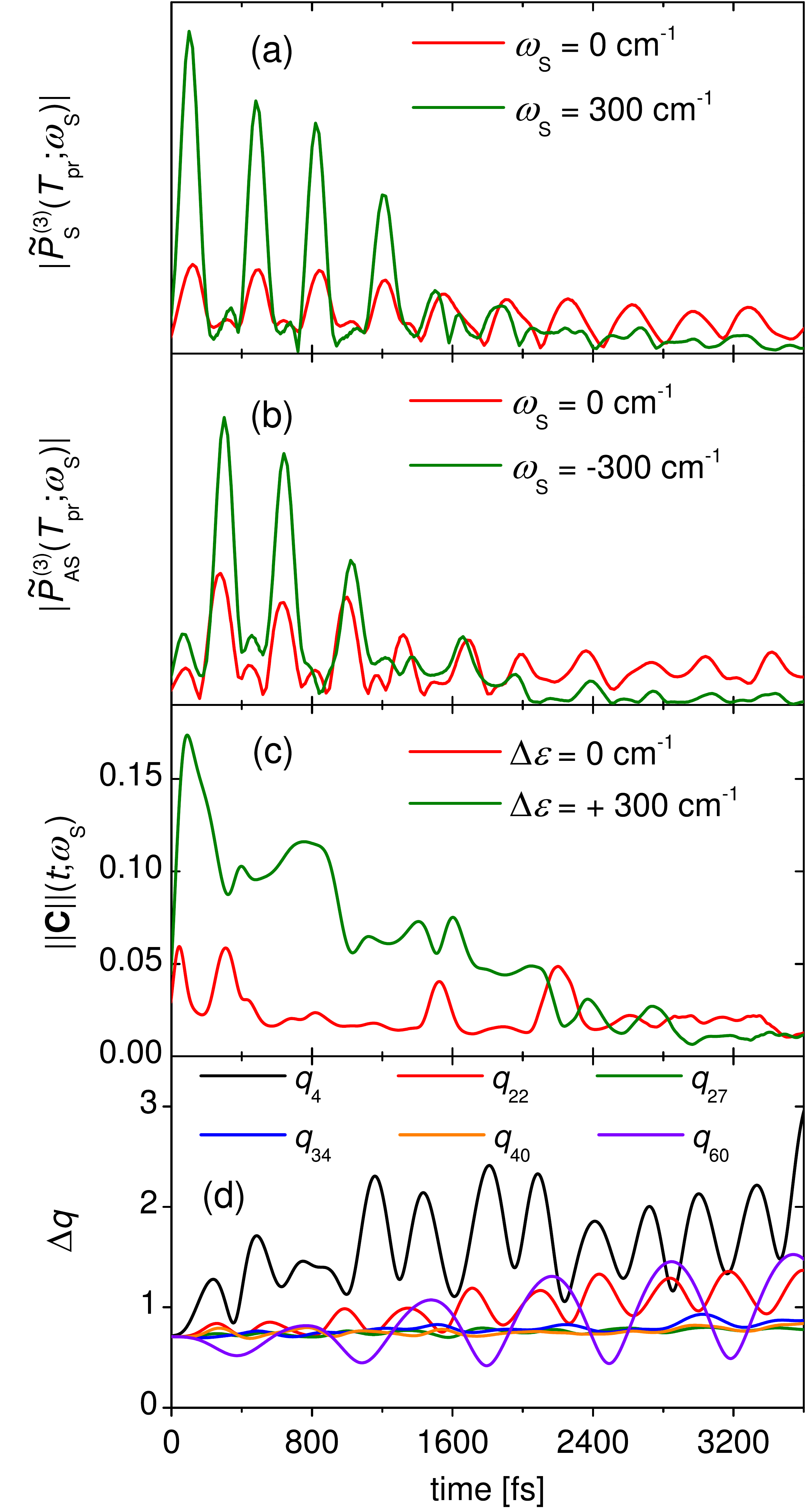}
\caption{(a,b)  One-dimensional cuts of the time-resolved Stokes and anti-Stokes coherent Raman signals $\left|\widetilde{P}_\mathrm{S}^{(3)}(T_\Pro)\right|$ and $\left|\widetilde{P}_\mathrm{AS}^{(3)}(T_\Pro)\right|$ of Fig. \ref{fig: spectrum BB} for fixed values of the Raman shift: $\omega_S = 0, 300\,\mathrm{cm}^{-1}$ for S and $\omega_\mathrm{S} = 0,-300\,\mathrm{cm}^{-1}$ for AS.  (c) The coherence norm $||\mathbf{C}||(t;\omega_\mathrm{S})$ as a function of time for the Raman shifts of panel (b). (d) The average wave packet width $\Delta q_i$ for the bath modes as a function of time.} 
\label{fig: superposition BB}
\end{figure}

A more direct comparison between the time-resolved Raman signal and the time-dependent coherence is provided by Fig. \ref{fig: superposition BB}. Panels (a) and (b) show cuts of the two-dimensional CSRS and CARS signals at selected Raman shifts, marked with horizontal lines in the spectra of Fig. \ref{fig: spectrum BB}. Such values of $\omega_\mathrm{S}$ correspond to different diagonals of the $\mathbf{C}$ matrix, which are traced in the coherence maps of the figure and can be associated with the norm
\begin{equation}
||\mathbf{C}||(t;\omega_\mathrm{S}) = \left( \sum_{jl} \left| C_{jl}^{(12)}(t) \right|^2 \ee^{-\alpha_E \left( \varepsilon_j - \varepsilon_l - \hbar \omega_\mathrm{S} \right)^2} \right)^\frac{1}{2} \ ,
\end{equation}
where $\alpha_E$ is a suitable broadening parameter.
The time-dependent coherence norms associated with the cuts of Fig. \ref{fig: spectrum BB} are shown in Fig. \ref{fig: superposition BB}(c). The clear-cut oscillations of the spectral peaks along the $T_\Pro$ axis are not present in the coherence norm traces; they result from the interference between the phase factors $\ee^{i(\varepsilon_j - \varepsilon_l)t/\hbar}$ in the sum of Eq. (\ref{eq: Spec2D_4}) and are refined by the convolution with the $\mathbf{W}^{\Omega_3,\Omega_\mathrm{R}}$ matrix. 

Nevertheless, several analogies are found between the plots of Figs. \ref{fig: superposition BB}(a,b) and (c).  The off-diagonal coherence decays with a multi-step mechanism. The first iodine-krypton collision leads to the initial decay that occurs between 100\,fs and 300\,fs and is associated with the intensity decrease between the first and second peak of the Stokes sequence. A second decrease is found between 800\,fs and 1000\,fs and is consistent with the dropoff simultaneously observed in the CSRS signal at $\omega_\mathrm{S} = 300\,\mathrm{cm}^{-1}$ and in the CARS signal at $\omega_\mathrm{S} = -300\,\mathrm{cm}^{-1}$. The subsequent decay steps, at 1600$-$1700\,fs, 2100$-$2200\,fs and 2750$-$2950\,fs, correlate well with the decrease of intensity observed in both the CSRS and in the CARS signals at $\omega_\mathrm{S} \ne 0$. The norm of the diagonal coherence is almost stationary as a function of time and only features few peaks which do not have a clear equivalent in the spectral cuts. A nearly constant amplitude of oscillation, in line with the behavior of the coherence, is found only in the CSRS trace at $\omega_\mathrm{S} = 0\,\mathrm{cm}^{-1}$. On the contrary, a step-like decline is still observed at 1200\,fs in the CARS cut at $\omega_\mathrm{S} = 0\,\mathrm{cm}^{-1}$; therefore the CARS signal is not only a genuine signature of the coherence dynamics, but is also strongly affected by the TFG convolution function.

The limited number of cage modes prevents an extensive dissipation, and the decoherence rate is also underestimated compared to the experimental conditions.\cite{SKFA05} Nevertheless, the calculated overall spectra and the selected traces are in remarkable agreement with the signals measured by Segale and Apkarian using similar values for pulse durations and energy separation between pulses.\cite{SA11}  A number of features found in the experiments are reproduced by the computations: (i) The variation of the Raman emission maxima and band width as a function of pump-probe delay; (ii) the delayed dropoff of the CSRS signal compared to the CARS signal (1.4\,ps vs 1.1\,ps), which is mostly a consequence of the fact that the coherence contributing to the latter one is partially unobserved due to the convolution with the TFG function of Eq. (\ref{eq: Spec2D_4}); (iii) the slippage of phase between the time oscillations in the CSRS and CARS cuts, which is recognizable both in Figs. \ref{fig: spectrum BB}(a,b) and \ref{fig: superposition BB}(a,b); (iv) the frequency doubling observed in the CARS and CSRS cuts at $\omega_\mathrm{S} = \pm 300\,\mathrm{cm}^{-1}$ and $T_\mathrm{pr} > 1.2\,\mathrm{ps}$, which emerges as a splitting of the vibrational peaks along the time axis. The origin of the features (iii) and (iv) was explained by Segale and Apkarian using a phase space picture of the moving wave packets.\cite{SA11} For short times the wave packets $\chi_1^B$ and $\chi_2^B$ have different energies. Due to the ultrafast electronic dephasing between the states $B$ and $E$, $\chi_2^B$ can be probed by the Stokes process only when it crosses the probe region moving towards the direction of I--I bond elongation; in contrast, in the anti-Stokes detection $\chi_2^B$ is probed when it moves towards bond compression, therefore with some delay compared to the Stokes case. After dissipation, the fact that the wave packets $\chi_1^B$ and $\chi_2^B$ get similar energies, allows the probe of $\chi_2^B$ either when it moves inwards and outwards, leading to a split of the Raman peaks similar to the one observed in pump-probe experiments.\cite{BGDS02,SA11}

In order to gain more insight into the dissipation and decoherence mechanism, the broadening of the wave packets is analyzed. The average variance of the wave packets $\chi_1^B$ and $\chi_2^B$ along the bath modes is calculated as
\begin{equation}
 \Delta q_i^2 = \frac{1}{2} \sum_{\alpha = 1,2} \left( \left\bra \chi_\alpha^B \left| q_i^2 \left| \chi_\alpha^B \right\ket \right. \right. - \left\bra \chi_\alpha^B \left| q_i \left| \chi_\alpha^B \right\ket \right. \right.^2 \right) \ .
\end{equation}
The $B \longleftarrow X$ excitation induces a prompt elongation of the I--I bond; for bath modes which are poorly correlated with the system mode, the wave packet width $\Delta q_i$ is expected to have little variations as time increases, even if the mode is strongly displaced upon $B \longleftarrow X$ excitation. In contrast, strong variations of the wave packet width are predictable for bath degrees of freedom correlated to the I--I motion, which are also expected to drive decoherence.  
Fig. \ref{fig: superposition BB}(d) reports the standard deviations $\Delta q_i$ and shows that the wave packet width variations are by far the strongest along the mode $q_4$, i. e. the stretch of the belt Kr atoms in the plane orthogonal to the I--I bond (see Fig. \ref{fig: modes}). In the first 800\,fs the bath modes are nearly inactive, but $\Delta q_4$ already starts growing and oscillating; this initial behavior leads to the first two steps of the off-diagonal coherence decay shown in Fig. \ref{fig: superposition BB}(c).  
 The relevance of the mode $q_4$ could have been anticipated by the preliminary classical trajectory study of Paper I.\cite{PCB18A} In a large scale model of the $\mathrm{I_2:Ar}$ system, a strong correlation between a cage mode similar to $q_4$ and the I--I stretch was found using classical dynamics simulations, which were indeed successful in reproducing most features of the experimental pump-probe spectra.\cite{LZAM95}

\begin{figure*}
\centering
\includegraphics[width=1.0\textwidth]{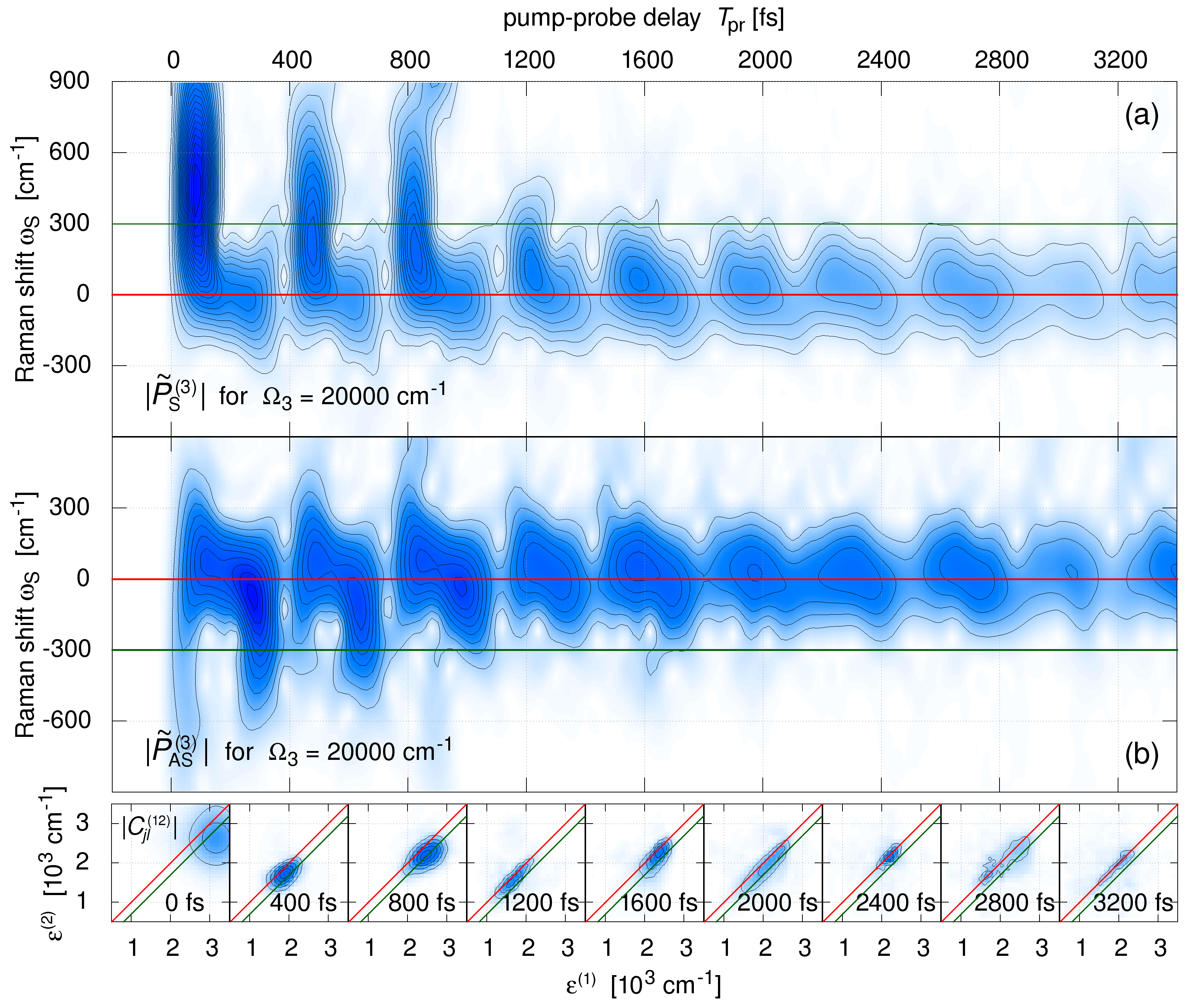}
\caption{The same as in Fig. \ref{fig: spectrum BB}, but for the initial wave packet superposition \textbf{B} (see Table \ref{tab: cat parameters}).}
\label{fig: spectrum D}
\end{figure*}

After 800\,fs the oscillations of $\Delta q_4$ become more regular and with larger amplitude, and two additional slower modes get activated. One of them is the cage breathing mode $q_{22}$ (see Fig. \ref{fig: modes}), whose oscillation amplitude grows stepwise every 750--800\,fs, in nice correlation with the multi-step decay of the coherence and with the decline of the CSRS and CARS signals. The other active mode, $q_{60}$, is the pistonlike translational motion of the whole $\mathrm{I_2}$ molecule in the krypton cavity; the width $\Delta q_{60}$ oscillates with increasing amplitude, probably as a consequence of the cage dilation along the mode $q_{22}$. The remaining modes, $q_{27}$, $q_{34}$ and $q_{40}$ behave as \lq spectators'.

\subsection{ Superposition \textbf{B}} 
In this case the wave packets $\chi_1^B$ and $\chi_2^B$ are prepared by $B \longleftarrow X$ excitations stimulated by laser pulses shorter than 20\,fs. The field parameters, reported in Table \ref{tab: cat parameters}, are adjusted to replicate the experiments performed in Ref. \onlinecite{SA11} using energetically broad pulses separated by $\sim 660\,\mathrm{cm}^{-1}$ [see Fig. \ref{fig: Pulses}(d)]. The time-resolved coherent Stokes and anti-Stokes Raman spectra for the superposition \textbf{B} are shown in Fig. \ref{fig: spectrum D}. The spectral cuts at $\omega_\mathrm{S} = 0, \pm300\,\mathrm{cm}^{-1}$ are marked on the spectra and the same is done for the corresponding lines in the $\mathbf{C}$ matrix.

 In the CSRS signal the first Raman band has an intense maximum that peaks at $\omega_\mathrm{S} \approx 450\,\mathrm{cm}^{-1}$. As expected from the higher spectral separation between the pulses (see Fig. \ref{fig: Pulses}), the Raman shift is higher than that of the earliest band maxima in the spectra of the superposition \textbf{A}. The value of $450\,\mathrm{cm}^{-1}$ is however rather low compared to the energy separation between the initially prepared wave packets, because of the ultrafast ($< 100\,\mathrm{fs}$) initial dissipation of the more energetic wave packet $\chi_1^B$. Similarly to the case \textbf{A}, the CSRS intensity declines in a stepwise fashion, with dropoffs at $T_\mathrm{pr} \approx 400\,\mathrm{fs}$ and $T_\mathrm{pr} \approx 1000\,\mathrm{fs}$. The emission maxima decrease rapidly first to $\omega_\mathrm{S}=170-220\,\mathrm{cm}^{-1}$ in the time range $400-1000\,\mathrm{fs}$, then to $\omega_\mathrm{S} \approx 80\,\mathrm{cm}^{-1}$ for $1100\,\mathrm{fs} < T_\mathrm{pr} < 1700\,\mathrm{fs}$, and finally towards the degenerate region; similarly, the emission band width decreases monotonically from $820\,\mathrm{cm}^{-1}$ and $T_\mathrm{pr} = 80\,\mathrm{fs}$ to $370\,\mathrm{cm}^{-1}$ at $T_\mathrm{pr} = 2\,\mathrm{ps}$.

The two-dimensional pattern of the CSRS spectrum faithfully describes  the dynamics of the wave packet coherence, reported in the bottom panels of Fig. \ref{fig: spectrum D}. The $\mathbf{C}$ matrix undergoes noticeable oscillations along the directions both parallel and perpendicular to the energy diagonal. 
As for the case \textbf{A}, the parallel oscillations indicate a non-monotonic dissipation, whereas the coherence dynamics along the anti-diagonal can be traced back in the spectrum by the motion of the emission maximum. Such visual inspection is facilitated in Fig. \ref{fig: spectrum D} by the red and green lines, and the Raman shift nicely matches the distance of the $\mathbf{C}$ matrix from the energy diagonal. Despite the fact that the initial coherence width is considerably larger than for the superposition \textbf{A} ($1220\,\mathrm{cm}^{-1}$ and $1120\,\mathrm{cm}^{-1}$ along the $\varepsilon^{(1)}$ and $\varepsilon^{(2)}$ axes, respectively), the $\mathbf{C}$ matrix shrinks immediately after the first I--I bond elongation, so that the frequency range of the S emission is the same in Figs. \ref{fig: spectrum BB} and \ref{fig: spectrum D}. In contrast to the case \textbf{A}, however, the coherence remains rather large and broad -- along both the energy diagonal and the anti-diagonal -- during the whole dynamics. 

As explained in Sect. \ref{sec: case A}, the CARS spectrum is a less accurate interpreter of the coherence dynamics, due to the larger filtering operated by the TFG matrix $\mathbf{W}^{\Omega_3,\Omega_\mathrm{R}}$.  In this case, even the Raman shift at the first emission maximum ($\omega_\mathrm{S} = - 75\,\mathrm{cm}^{-1}$) is very different from the initial wave packet energy separation ($\sim 660\,\mathrm{cm}^{-1}$), and the FWHM of the first band ($\approx 400\,\mathrm{cm}^{-1}$) is much smaller than the coherence width. In addition, the facts that the signal centralizes in the first picosecond, the intensity decays extremely slowly and the width remains nearly constant, would incorrectly suggest that the (anti-diagonal) coherence motion rapidly becomes stationary.  The shape of the earliest two-dimensional peaks and the overall behavior of the signal nicely agree with the  experimental spectrum reported in Fig. 12 of Ref. \onlinecite{SA11}  (the CSRS spectrum is not reported in the paper), where the authors indeed attribute the long-lived sequence of peaks to a strong slowdown of the dissipation after the first chromophore-cage interaction.\cite{note_PCB18B} The present simulations partially contradict this picture and show that the absence of dynamical features in the time-resolved spectrum does not necessarily imply an arrested coherence dynamics. The fact that the probe pulse provides a filtered picture of the wave packet coherence implies that, in order to fully map the Schr\"{o}dinger cat dynamics, it is highly beneficial to perform measurements using a number of different field parameters.

\begin{figure}[b!]
\centering
\includegraphics[scale = 0.3]{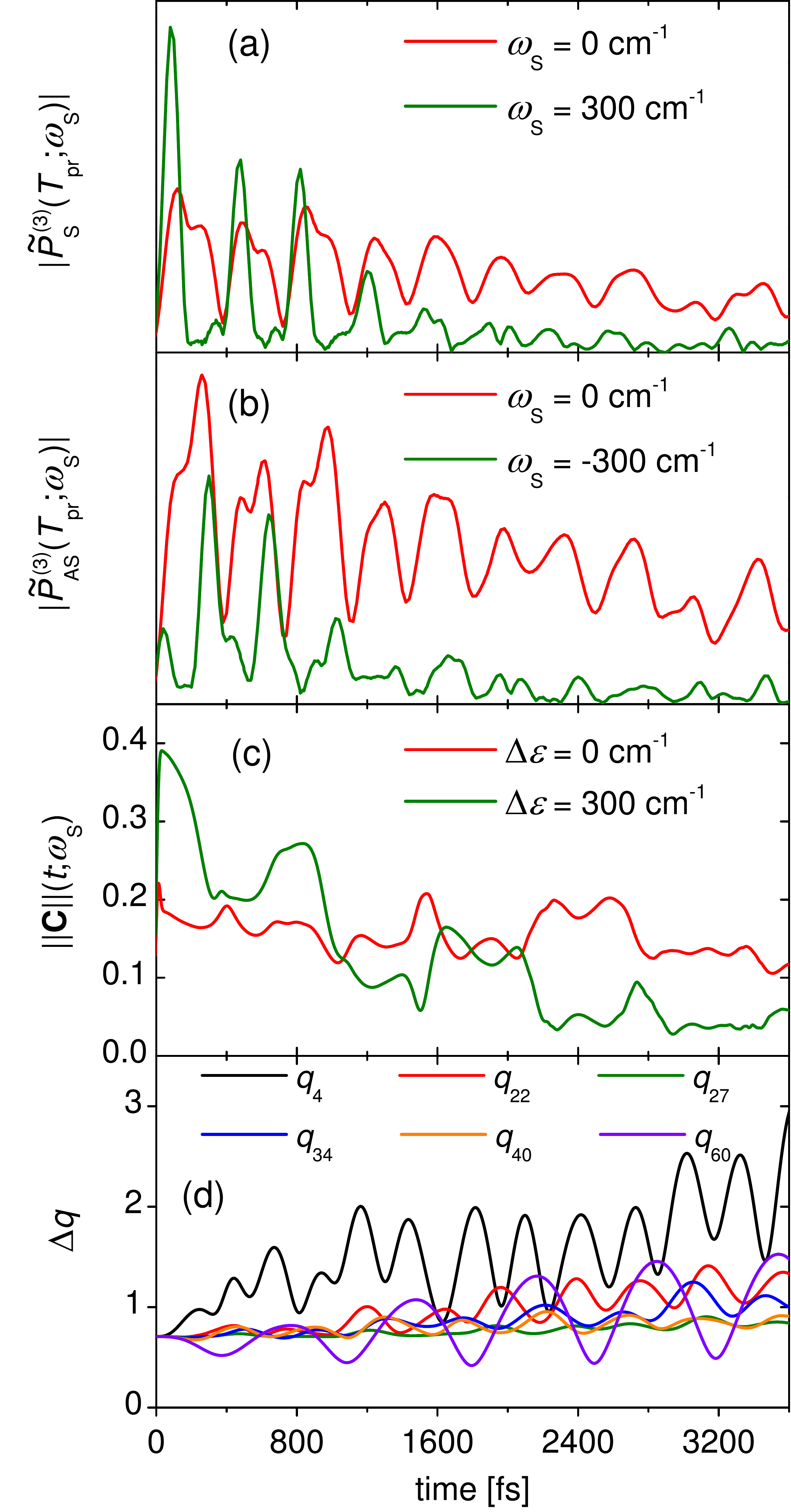}
\caption{(a,b)  One-dimensional cuts of the time-resolved Stokes and anti-Stokes coherent Raman signals $\left|\widetilde{P}_\mathrm{S}^{(3)}(T_\Pro)\right|$ and $\left|\widetilde{P}_\mathrm{AS}^{(3)}(T_\Pro)\right|$ of Fig. \ref{fig: spectrum D} for fixed values of the Raman shift: $\omega_S = 0, 300\,\mathrm{cm}^{-1}$ for S and $\omega_\mathrm{S} = 0,-300\,\mathrm{cm}^{-1}$ for AS.  (c) The coherence norm $||\mathbf{C}||(t;\omega_\mathrm{S})$ as a function of time for the Raman shifts of panel (b). (d) The average wave packet width $\Delta q_i$ for the bath modes as a function of time.} 
\label{fig: superposition D}
\end{figure}

Spectral cuts at selected Raman frequencies and the corresponding coherence norms are shown in Fig. \ref{fig: superposition D}(a), (b) and (c). Differently from the case \textbf{A}, the long lived diagonal coherence is manifested in the nearly undamped oscillations of the CSRS and CARS signals for $\omega_\mathrm{S} = 0\,\mathrm{cm}^{-1}$. On the other hand, the cuts at $\pm\omega_\mathrm{S} = 300\,\mathrm{cm}^{-1}$ behave similarly to the ones of Fig. \ref{fig: superposition BB}; in particular, decay steps at the same times (1.2\,ps and 2\,ps) are observed. Such features are traced back to the stepwise decay of the off-diagonal coherence norm, shown in Fig. \ref{fig: superposition D}(c); the lack of decay in the diagonal coherence is instead consistent with the long lived signals at $\omega_\mathrm{S} = 0\,\mathrm{cm}^{-1}$. Compared to the superposition \textbf{A}, the coherence norm behaves similarly but undergoes more \lq noisy' oscillations, due to the fact that the broader $\mathbf{C}$ matrix captures bath dynamics over a larger energy range; moreover, comparing the norms for $\varepsilon^{(1)} = \varepsilon^{(2)}$ and $\varepsilon^{(1)} = \varepsilon^{(2)} + 300\,\mathrm{cm}^{-1}$ in Figs. \ref{fig: superposition BB}(c) and \ref{fig: superposition D}(c), one concludes that the preparation \textbf{B} induces a larger diagonal coherence norm, which is the reason for the long lived peak sequences in the spectra.

The mechanistic events which drive dissipation and decoherence are made clear by the analysis of the average wave packet width $\Delta q_i$ along bath modes, which is depicted in Fig. \ref{fig: superposition D}(d). The variations of $\Delta q_i$ as a function of time resemble the ones of Fig. \ref{fig: superposition BB}(c), suggesting that similar dissipation mechanisms are operative in the preparations \textbf{A} and \textbf{B}. Also in this case, the decoherence steps correlate with the transfer of energy from the I--I vibration to the surroundings, which give rise to wave packet broadening along bath modes: The simultaneous decrease of the off-diagonal coherence norm and the CSRS and CARS signals around 1.2\,ps correlates well with the growth of $\Delta q_4$ and $\Delta q_{22}$ (i. e. with the belt atoms stretch and Kr cavity breathing mode); in correspondence with the decline at 2\,ps, the bath wave packet broadens mostly along modes $q_{22}$ and $q_{60}$ (pistonlike translational motion of $\mathrm{I_2}$); finally, the decoherence step around 2.8\,ps is most likely driven by the modes $q_4$, $q_{22}$ and $q_{34}$ (belt atoms stretch). Compared to case \textbf{A}, where most bath modes were passive, the dynamics following the preparation \textbf{B} involves the activity of a larger number of modes. The dominant role in dissipation and decoherence is always played by the modes $q_4$, $q_{22}$ and $q_{60}$, but $q_{34}$ and $q_{40}$ are also operative, albeit to a lesser extent.


\section{Conclusion}
\label{sec: Conclusion}

The G-MCTDH method is a cheap and powerful means to investigate the photodynamics of chromophores embedded in a matrix. In particular, the present study shows that G-MCTDH simulations allow to understand which signatures of the dissipation and decoherence dynamics are imprinted in
nonlinear optical spectra. 
The method allows the efficient calculation of time-resolved coherent Stokes and anti-Stokes Raman signals of the iodine chromophore in solid krypton. The spectra, calculated for specific Schr\"{o}dinger cat wave packet superpositions created by $B \longleftarrow X$ transitions, have
many features in common with the experimental spectra, as for example the
period of oscillation as a function of the pump-probe delay, the shrinking and the convergence of the Raman bands at long times. Also nicely reproduced is the stepwise decay of the Raman intensity observed in some measurements,\cite{SA11} as a consequence of the various time scales of the dynamical events which drive decoherence. The motion of the krypton cage modes responsible for such  mechanism is visualized in detail in the G-MCTDH simulations.

The spectral features can be related to the underlying molecular dynamics using
the theoretical treatment developed in Sect. \ref{sec: theory}. The signal
is interpreted as the trace of the convolution between two matrices: (i) The chromophore vibrational coherence matrix,
which describes the correlation between the wave packet pair and depends only on the $B \longleftarrow X$ pump excitation process; (ii) a time and frequency gate matrix function, which depends on the probe pulse specifics and on Franck-Condon factors,
therefore it embeds information about the topography of the potential energy
surfaces. 
The motion of the reduced coherence is therefore not observed directly, but is filtered by the probe pulse. As proved by the present simulations, a single Stokes or anti-Stokes signal might be insufficient to fully visualize the Schr\"{o}dinger cat dynamics, or can even lead to a misleading interpretation of the process. It is instead advisable to detect the Raman signal simultaneously in the S and AS directions, preferably using probe pulses of different frequency and duration, as suggested by the previous experiments of Apkarian and coworkers.\cite{SKFA05,SA11}

Nevertheless, dynamical features in the four-wave-mixing spectra, like sudden signal dropoffs, are not spurious, but are related to the main molecular motions which drive the energy exchange between the guest chromophore and the crystal host.
The comparison between the cases \textbf{A} and \textbf{B} illustrates the fact
that the dissipation mechanism is not completely \lq universal'. The
stretching mode of the belt Kr atoms (mode $q_4$) is always operative immediately during the first I--I bond elongation; at slightly longer times ($\sim 1\,\mathrm{ps}$) the cage breathing mode $q_{22}$ takes energy from the guest, and the oscillation amplitude of the pistonlike motion $q_{60}$ grows steadily, indicating an increasing freedom of movement of the iodine molecule in the krypton cavity. Thus, $q_{22}$ and $q_{60}$ can be also regarded as important dissipative modes. 
The same cannot be stated for the other modes, whose
involvement in the decoherence dynamics depends on the initial wave packets'
preparation. The number of active bath modes is larger in the superposition
\textbf{B} than in case \textbf{A}, in which the decoherence is actually slightly 
faster and the Raman signal decline more rapidly. Indeed, the strong anharmonicities give rise to non-trivial
system-bath interactions, by which the activity of a larger number of
dissipative modes does not necessarily imply faster decoherence.

As a final remark, the approximate Eq. (\ref{eq: Spec2D_4}) for the nonlinear spectrum is
found to be useful not only for interpretative but also for computational
purposes. The time-resolved coherent Raman signal can be calculated with a
minimum number of wave packet propagations (four) and the time-dependent
wavefunctions do not need to be saved in memory. This is especially useful in
future simulations of four-wave-mixing experiments for larger clusters, which will include more coordinates and, possibly, non-adiabatically coupled electronic states.\cite{BC99} In these cases, the number of spectra obtained with different carrier frequencies, time delays or
pulse durations is large, and the efficient spectral simulations allow to
inspect the signals for several of such setups. The combinations of field
parameters which are most informative about the mechanisms of decoherence and
dissipation can be identified, and the calculation of the corresponding time-dependent CARS and CSRS spectra can be improved using the accurate Eq. (\ref{eq: P3 expand}).

\section*{Acknowledgments}
J. A. C. acknowledges the support given by the US-NSF Grant No. CHE 1565680.

\appendix

\section{Comparison between the spectra calculated using Eqs. (\ref{eq: P3 expand}) and (\ref{eq: Spec2D_4})}
In this Appendix, the time-resolved CSRS and CARS spectra, calculated using the accurate Eq. (\ref{eq: P3 expand}) and the approximate Eq. (\ref{eq: Spec2D_4}), are compared. Eq. (\ref{eq: Spec2D_4}) is based on the approximation of Eq. (\ref{eq: C approx}), and is computationally more advantageous because the coherence matrix $\widetilde{C}_{jl}^{(12)}(t) = \left\bra \chi_1^B,t \left| \phi_j \right\ket  \right.  \left\bra \phi_l \left| \chi_2^B,t \right\ket \right.$ can be evaluated without explicitly saving the full time-dependent wave packets $\chi_1^B(t)$ and $\chi_2^B(t)$. To this purpose, quantum dynamical calculations can be performed on the $B$ state surface for four auxiliary vibrational wavefunctions, 
\begin{equation}
|\Psi_\alpha^B,t\ket = \left| \chi_2^B,t \right\ket + \ee^{i\alpha} \left| \chi_1^B,t\right\ket \ ,  \mbox{ with } \alpha = 0,\frac{\pi}{2},\pi,-\frac{\pi}{2} \ ,  \label{eq: psi_alpha}
\end{equation}
and the only quantities which must be saved in memory during the propagation runs are the reduced density matrices in the energy representation,
\begin{equation}
 \varrho_{\alpha,jl} = \bra \Psi_\alpha^B,t | \phi_j \ket \bra \phi_l | \Psi_\alpha^B,t \ket \ .
\end{equation}
 Note that the equations of motions for the four wavefunctions can be integrated separately, with the advantage of having integration step sizes adapted to the individual propagations. 

The coherence matrices can be finally obtained as
\begin{eqnarray}
 \varrho_{0,jl} - \varrho_{\pi,jl} + i \varrho_{\frac{\pi}{2},jl} - i\varrho_{-\frac{\pi}{2},jl} & = & \nonumber \\ 
  & & \hspace{-2.5cm} + \left\bra \chi_2^B+\chi_1^B \left| \phi_j  \right\ket \right. \left\bra \phi_l \left| \chi_2^B + \chi_1^B \right\ket \right. \nonumber \\
 &  & \hspace{-2.5cm}  - \left\bra \chi_2^B - \chi_1^B \left| \phi_j  \right\ket \right. \left\bra \phi_l \left| \chi_2^B - \chi_1^B \right\ket \right.  \nonumber \\
 &  & \hspace{-2.5cm}  +i \left\bra \chi_2^B + i\chi_1^B \left| \phi_j \right\ket \right. \left\bra \phi_l \left| \chi_2^B + i\chi_1^B \right\ket \right.  \nonumber \\
 &  & \hspace{-2.5cm}  -i \left\bra \chi_2^B - i\chi_1^B \left| \phi_j \right\ket \right.  \left\bra \phi_l \left| \chi_2^B - i\chi_1^B \right\ket \right.  \nonumber  \\  
 &  &  \hspace{-2.5cm}  = \left\bra \chi_2^B \left| \phi_j  \right\ket \right. \left\bra \phi_l \left| \chi_2^B \right\ket \right. (1 - 1 + i - i) \nonumber \\
 &   & \hspace{-2.5cm}  + \left\bra \chi_2^B \left| \phi_j \right\ket \right. \left\bra \phi_l \left| \chi_1^B \right\ket \right. (1 + 1 - 1 - 1) \nonumber \\
 &   & \hspace{-2.5cm}  + \left\bra \chi_1 \left| \phi_j \right\ket \right. \left\bra \phi_l \left| \chi_2 \right\ket \right. (1 + 1 + 1 + 1) \nonumber \\
 &   & \hspace{-2.5cm}  + \left\bra \chi_1 \left| \phi_j \right\ket \right. \left\bra \phi_l \left| \Psi_1 \right\ket \right. (1 - 1 + i - i) \nonumber \\
 &  &  \hspace{-2.5cm}  =  4 \widetilde{C}_{jl}^{(12)}(t) \ .
\end{eqnarray}

Auxiliary wavefunctions different from the ones of Eq. (\ref{eq: psi_alpha}) can also be used. However, since only one off-diagonal coherence component $ \widetilde{C}_{jl}^{(12)}(t)$ [and not $  \widetilde{C}_{jl}^{(12)}(t)  + \widetilde{C}_{jl}^{(21)}(t) $]  needs to be calculated, the number of wavefunctions which need to be propagated -- without storing them in memory -- cannot be reduced to less than four. 

\begin{figure}[b!]
\centering
\includegraphics[scale=0.095]{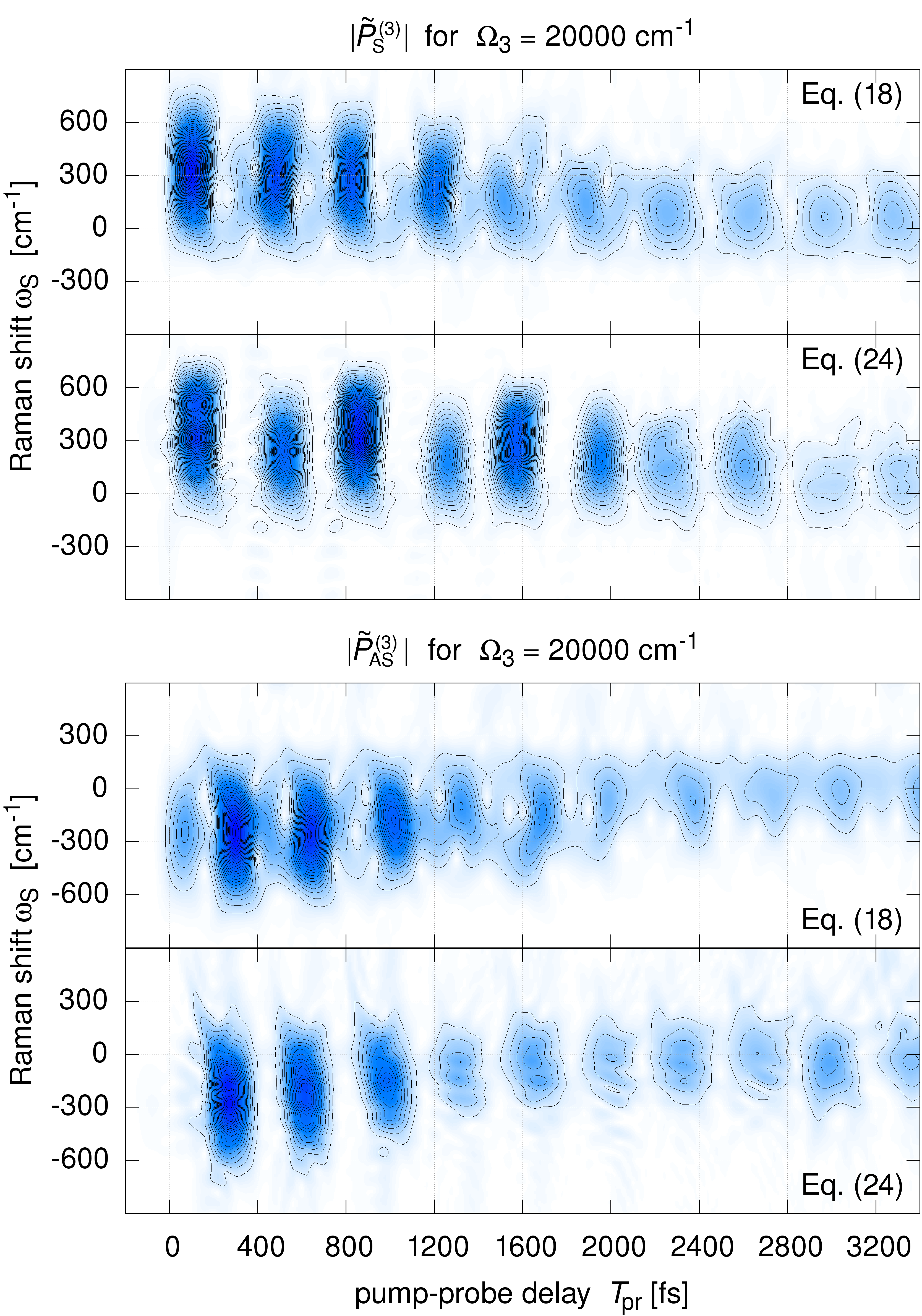}
\caption{Comparison between the time-resolved CSRS (top panels) and CARS (bottom panels) spectra calculated using Eq. (\ref{eq: P3 expand}) vs. Eq. (\ref{eq: Spec2D_4}) for the wave packet superposition \textbf{A}, as a function of pump-probe delay and Raman shift $\omega_\mathrm{S} = \Omega_3 - \Omega_\mathrm{R}$. The probe central frequency is set to $\Omega_3 = 20000\,\mathrm{cm}^{-1}$. Eq. (\ref{eq: P3 expand}) is based on the full two-times cross-correlation matrix $C_{jl}^{(12)}(t,\tau)$, whereas Eq. (\ref{eq: Spec2D_4}) exploits the approximation $C_{jl}^{(12)}(t,\tau) \approx C_{jl}^{(12)}(t^\prime,t^\prime)$, with $t^\prime = (t + \tau) / 2 $}
\label{fig: comp A}
\end{figure}

The CSRS and CARS spectra obtained using Eqs. (\ref{eq: P3 expand}) and (\ref{eq: Spec2D_4}) are compared for the Schr\"{o}dinger cat superposition \textbf{A} in Fig. \ref{fig: comp A}. The result of the approximation formula nicely captures the recurrence period along the $T_\mathrm{pr}$-axis, the shift of of the emission maxima from $|\omega_\mathrm{S}| \approx 300\,\mathrm{cm}^{-1}$ to the long time value of $\omega_S \approx 0\,\mathrm{cm}^{-1}$, and the decoherence events which lead to the stepwise decrease of the emission intensity, as discussed in Sect. \ref{sec: case A}. For the CARS signal, the agreement is nearly perfect. For the Stokes signal, the widths and positions of the emission bands are well reproduced, but in the approximation of Eq. (\ref{eq: Spec2D_4}) the intensity of the first five bands oscillates instead of decreasing monotonically. The worse performance of Eq. (\ref{eq: Spec2D_4}) for the Stokes case suggests that the approximation based on Eq. (\ref{eq: C approx}) is less accurate when the emission terminates in the vibrational levels having high energy, like in the CSRS process (see Fig. \ref{fig: Pulses}). This is consistent with the fact that for large pump-probe delays, i. e. after partial vibrational relaxation, both the CSRS and CARS spectra are correctly reproduced.

\begin{figure}[b!]
\centering
\includegraphics[scale=0.095]{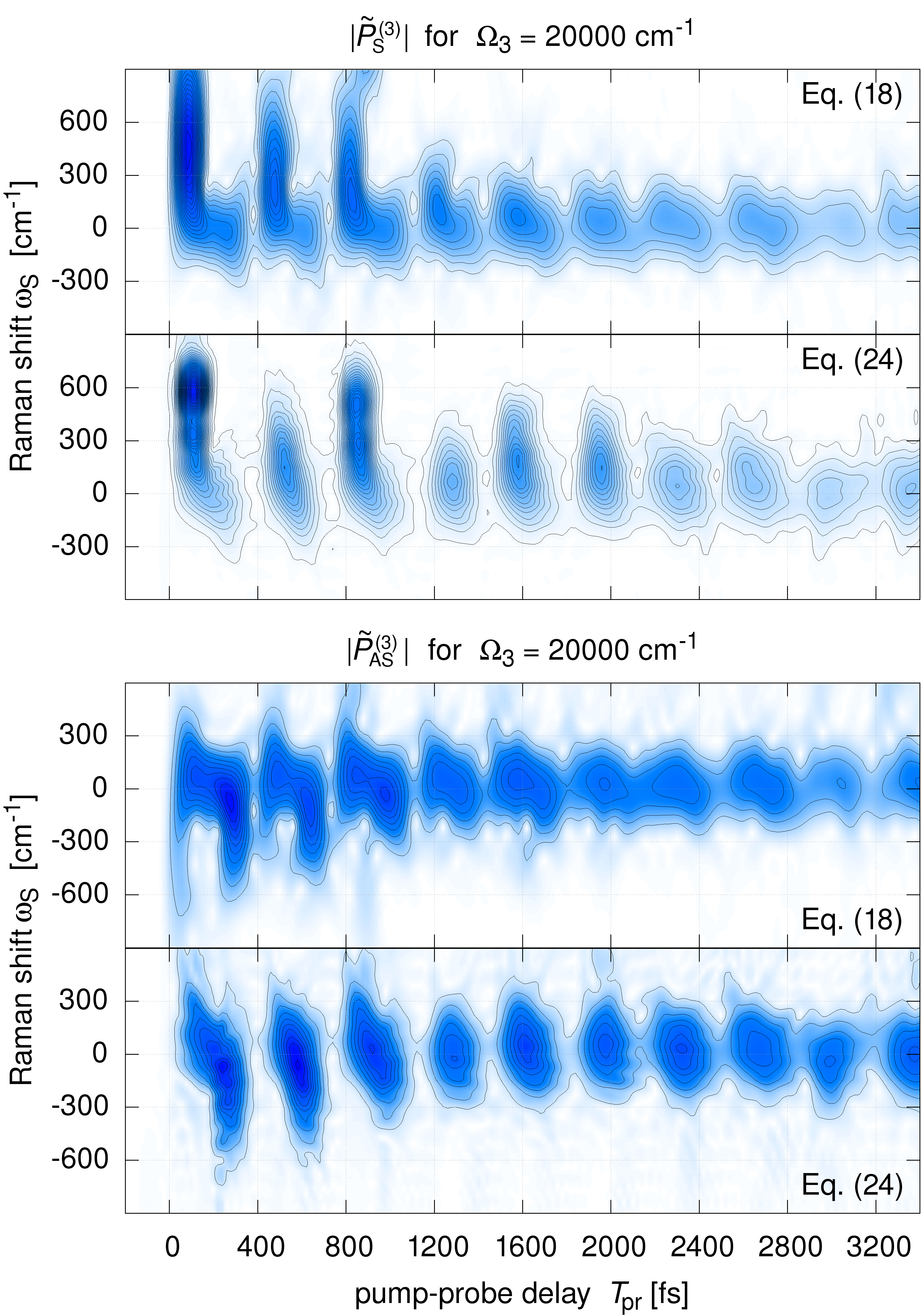}
\caption{The same as in Fig. \ref{fig: comp B}, but for the wave packet superposition \textbf{B}.}
\label{fig: comp B}
\end{figure}

For the superposition \textbf{B}, the comparison between the spectra calculated using the exact and the approximate cross-correlation matrix is illustrated in Fig. \ref{fig: comp B}. Also in this case the approximate CSRS and CARS spectra agree with the accurate ones. The decay steps and the dependence of the emission maxima on the pump-probe delay are nicely recovered. The approximation of the CARS signal correctly predicts the $\approx 300\,\mathrm{cm}^{-1}$ discrepancy between the earliest Raman shift and the energy separation between the initial wave packets. The time-dependent emission band width and the slow decay of the signals are also well reproduced by the approximate expression of Eq. (\ref{eq: Spec2D_4}). Similarly to the case \textbf{B}, the approximation of Eq. (\ref{eq: C approx}) works worse for the Stokes spectrum and the peak intensities decay in an oscillatory fashion instead than monotonically. Nevertheless, the dropoff at $T_\mathrm{pr} \approx 1.2\,\mathrm{ps}$ is nicely predicted, and the behavior of the signal at large values of $T_\mathrm{pr}$ is described correctly. These findings agree with the conjecture that the approximation is less good when the arrival wave packet has components on a number of high lying vibrational levels. 

On the whole, the estimate of the time- and frequency-dependent spectral shape obtained from Eq. (\ref{eq: Spec2D_4}) is good. Given that this equation does not require the storage of the wavefunction and the evaluation of wave packet overlaps for many time pairs, it can be used to investigate a large set of combinations of field parameters. The results of major interest in this set can be studied in more detail and the spectra can be improved using the formula of Eq. (\ref{eq: P3 expand}).


\begin{thebibliography}{43}
\bibitem [{not()}]{note_PCB18B}%
  \BibitemOpen
  \href@noop {} {}\bibinfo {note} {The experimental spectrum contains a band at
  $T_\mathrm{pr} \approx 0$ with a Raman shift $\omega_\mathrm{S} \approx
  -600\,\mathrm{cm}^{-1}$. This feature is probably due to additional
  processes, described by Feynman pathways which are non-vanishing only at very
  short times ($T_\mathrm{pr} \lesssim 150\,\mathrm{fs}$) and which are not
  included in the simulations of this work.}\BibitemShut {Stop}%
\end{thebibliography}

%

\end{document}